\documentclass[11pt,onecolumn,draftclsnofoot]{IEEEtran}
\usepackage{amsmath,amssymb,textcomp}
\usepackage{bbm}
\usepackage{mcode}
\usepackage{graphicx}
\usepackage{algorithm,algorithmic}
\newcommand{\LB}{\left(}
\newcommand{\RB}{\right)}
\newcommand{\LSB}{\left[}
\newcommand{\RSB}{\right]}
\newcommand{\htp}{^{\sf H}}
\newcommand{\tp}{^{\sf T}}

\newfont{\bbb}{msbm10 scaled 500}

\newfont{\bb}{msbm10 scaled 1100}
\newcommand{\CC}{\mbox{\bb C}}
\newcommand{\RR}{\mbox{\bb R}}

\newcommand{\EE}{\mbox{\bb E}}

\newcommand{\hv}{{\bf h}}

\newcommand{\mv}{{\bf m}}
\newcommand{\nv}{{\bf n}}

\newcommand{\wv}{{\bf w}}
\newcommand{\vv}{{\bf v}}
\newcommand{\xv}{{\bf x}}
\newcommand{\yv}{{\bf y}}
\newcommand{\zv}{{\bf z}}
\newcommand{\zerov}{{\bf 0}}

\newcommand{\Am}{{\bf A}}
\newcommand{\Bm}{{\bf B}}

\newcommand{\Dm}{{\bf D}}

\newcommand{\Gm}{{\bf G}}
\newcommand{\Hm}{{\bf H}}
\newcommand{\Id}{{\bf I}}

\newcommand{\Pm}{{\bf P}}
\newcommand{\Qm}{{\bf Q}}
\newcommand{\Rm}{{\bf R}}
\newcommand{\Sm}{{\bf S}}
\newcommand{\Tm}{{\bf T}}
\newcommand{\Um}{{\bf U}}
\newcommand{\Wm}{{\bf W}}

\newcommand{\Xm}{{\bf X}}

\newcommand{\Zm}{{\bf Z}}

\newcommand{\Cc}{{\cal C}}

\newcommand{\Fc}{{\cal F}}

\newcommand{\Ic}{{\cal I}}
\newcommand{\Jc}{{\cal J}}

\newcommand{\Nc}{{\cal N}}

\newcommand{\betav}{\hbox{\boldmath$\beta$}}

\newcommand{\Deltam}{\hbox{\boldmath$\Delta$}}

\newcommand{\diag}{{\hbox{diag}}}

\renewcommand{\det}{{\hbox{det}}}
\newcommand{\trace}{{\hbox{tr}\,}}

\renewcommand{\arg}{{\hbox{arg}}}

\newcommand{\defines}{{\,\,\stackrel{\scriptscriptstyle \bigtriangleup}{=}\,\,}}

\newtheorem{theorem}{Theorem}%[section]
\newtheorem{definition}{Definition}%[section]
\newtheorem{lemma}{Lemma}%[section]
\newtheorem{cor}{Corollary}

\newtheorem{remark}{Remark}[section] 
\newtheorem{example}{Example}
\newtheorem{property}{Property}
\newtheorem{assumption}{\indent \bf A}

\begin{document}
\bibliographystyle{IEEEtran}

\title{Iterative Deterministic Equivalents for the Performance Analysis of Communication Systems}
\author{Jakob~Hoydis$^{\star,\S}$, Romain~Couillet$^{\star}$ and M\'erouane~Debbah$^\star$
\thanks{Parts of this work have been presented at the Asilomar Conference on Signals, Systems, and Computers, CA, US, Nov. 2011.}
\thanks{$^\star$Alcatel Lucent Chair on Flexible Radio, Sup\'elec, 3 rue Joliot Curie, 91192 Gif sur Yvette, France.}
\thanks{$^\S$Department of Telecommunications, Sup\'elec, 3 rue Joliot Curie, 91192 Gif sur Yvette, France. }
\thanks{\{\scriptsize\texttt{jakob.hoydis,romain.couillet,merouane.debbah\}@supelec.fr}}}

\maketitle

\begin{abstract}
In this article, we introduce iterative deterministic equivalents as a novel technique for the performance analysis of communication systems whose channels are modeled by complex combinations of independent random matrices. This technique extends the deterministic equivalent approach for the study of functionals of large random matrices to a broader class of random matrix models which naturally arise as channel models in wireless communications. We present two specific applications: First, we consider a multi-hop amplify-and-forward (AF) MIMO relay channel with noise at each stage and derive deterministic approximations of the mutual information after the $K$th hop.  Second, we study a MIMO multiple access channel (MAC) where the channel between each transmitter and the receiver is represented by the double-scattering channel model. We provide deterministic approximations of the mutual information, the signal-to-interference-plus-noise ratio (SINR) and sum-rate with minimum-mean-square-error (MMSE) detection and derive the asymptotically optimal precoding matrices. In both scenarios, the approximations can be computed by simple and provably converging fixed-point algorithms and are shown to be almost surely tight in the limit when the number of antennas at each node grows infinitely large. Simulations suggest that the approximations are accurate for realistic system dimensions. 
The technique of iterative deterministic equivalents can be easily extended to other channel models of interest and is, therefore, also a new contribution to the field of random matrix theory.
\end{abstract}

\section{Introduction}
Since the pioneering work of Tse and Hanly \cite{TSE99} on the capacity of code division multiple access (CDMA) technologies assuming long spreading sequences, the theory of large dimensional random matrices (RMT) has drawn an increasing interest from researchers in wireless communications and related fields \cite{TUL04b,COUbook}. RMT is in particular convenient for the study of multiple-input multiple-output (MIMO) channels \cite{TEL99,HAC07}, (random) linear precoders \cite{TSE99,DEB03b,couillethoydis11}, multi-user systems \cite{NGU08,WAG10}, multi-cellular systems \cite{ZAK10,huh2010,HOY10}, etc. In the early contributions, it was systematically assumed that the dimensions of the system under study could grow infinitely large and that the system performance admitted a deterministic limit that RMT can provide \cite{TSE99,TEL99,TUL04}. It then became rapidly clear that, for most systems of practical interest, either the former assumption is not natural or the latter condition is not met. However, even for systems of large but finite size, the inherently random performance (e.g.\@ instantaneous mutual information, signal-to-interference-plus-noise ratio (SINR)), can often be well approximated by deterministic quantities. Such quantities are called \emph{deterministic equivalents}, and can be derived by various techniques, such as the Stieltjes transform method \cite{HAC07,COU09}, the Gaussian method \cite{DUP10,HAC06}, or the replica method \cite{MOU05,MOU07}.

Deterministic equivalents are convenient to study the performance of wireless communication systems when a single system parameter can be modeled by a random matrix, e.g. the fading channel or the spreading codes. In order to tackle the performance analysis of more complex systems, such as multi-hop communications, random beamforming over random fading channels, or double-scattering channels, it is necessary to extend the notion of deterministic equivalents. In the present article, which is inspired by the original idea of \cite{couillethoydis11}, where the performance of random isometric precoders over random fading channels is analyzed, we develop a systematic approach to generalize deterministic equivalents to {\it iterative deterministic equivalents}. To this end, we introduce a generic definition of deterministic equivalents of functionals of random matrices, which we extend, based on the Fubini theorem \cite{BIL08}, to a definition of iterative deterministic equivalents.

As application examples, we then provide deterministic equivalents of the mutual information of the multi-hop amplify-and-forward (AF) MIMO relay channel \cite{BOR07,MAR10b,FAW09} (Section~\ref{sec:relay}) and of the ergodic capacity as well as the sum-rate with minimum-mean-square-error (MMSE) detection of double-scattering multiple access channels (MACs) \cite{gesbert02,muller01} (Section~\ref{sec:double}). An overview of related research to both topics is provided in the respective sections. Our analysis is based on the Stieltjes transform method, documented in detail in \cite{COUbook}. 

The remainder of this article is structured as follows. In Section~\ref{sec:deteqdeteq}, we recall the fundamentals of deterministic equivalents in RMT and develop the notion of iterative deterministic equivalents. In Section~\ref{sec:applications}, we study applications of iterative deterministic equivalents to the performance analysis of  multi-hop relay channels and double-scattering MACs. The paper is concluded with Section~\ref{sec:conclusion}. All proofs, related results, and some exemplary Matlab codes are provided in the appendices.\bigskip

{\it Notations:} Boldface lower and upper case symbols represent vectors and matrices, respectively. $\Id_N$ is the size-$N$ identity matrix and $\diag(x_1,\dots,x_N)$ is a diagonal matrix with elements $x_i$. The trace, transpose, and Hermitian transpose operators are denoted by $\trace(\cdot)$, $(\cdot)\tp$, and $(\cdot)\htp$, respectively. The spectral norm of a matrix $\Am$ is denoted by $\lVert\Am\rVert$, and, for two matrices $\Am$ and $\Bm$, the notation $\Am\succ\Bm$ means that $\Am-\Bm$ is positive-definite. For a vector $\xv=[x_1\dots x_N]\tp$, $\xv\ge 0$ denotes $x_i\ge 0$ for all $i$. The notations $\Rightarrow$ and $\xrightarrow[]{\text{a.s.}}$ denote weak and almost sure convergence, respectively. We use $\Cc\Nc\left(\mv,\Rm\right)$ to denote the circular symmetric complex Gaussian distribution with mean $\mv$ and covariance matrix $\Rm$. We denote by $\RR^+$ the set $[0,\infty)$, by $\RR^-$ the set $(-\infty,0]$, and by ${\bf i} = \sqrt{-1}$. $\mathbbm{1}_A(x)$ is the indicator function, i.e., $\mathbbm{1}_A(x)=1$ iff $x\in A$ and $\mathbbm{1}_A(x)=0$ otherwise. $\mathbb{E}\LSB \cdot\RSB$ denotes the expectation operator. For $(a_n)_{n\ge1}$ and $(b_n)_{n\ge 1}$ two sequences of random variables, we denote $a_n\asymp b_n$ the equivalence relation $a_n-b_n\xrightarrow[]{\text{a.s.}} 0$ for $n\to\infty$.

\section{Iterative deterministic equivalents}
\label{sec:deteqdeteq}
In this section, we will first recall the notion of deterministic equivalents in probability theory before we explain their connections to RMT and the performance analysis of communication systems. We will then introduce the Fubini theorem, which is the key ingredient to extend classical deterministic equivalents to iterative deterministic equivalents.

\subsection{Deterministic equivalents and random matrices}
\begin{definition}
	\label{def:deteq}
Consider the probability space $(\Omega,\mathcal F,P)$. Let $(f_n)_{n\geq 1}$ be a series of measurable complex-valued functions, $f_n:\Omega\times \CC \to \CC$, and let $(g_n)_{n\geq 1}$ be a series of complex-valued functions, $g_n:\CC\to \CC$. Then $(g_n)_{n\geq 1}$ is a {\it deterministic equivalent} of $(f_n)_{n\geq 1}$ on $D\subset \CC$, if there exists a set $A\subset \Omega$ with $P(A)=1$, such that
	\begin{equation*}
f_n(\omega,z)-g_n(z) \xrightarrow[n\to\infty]{} 0
	\end{equation*}
	for all $\omega\in A$ and for all $z\in D$.
\end{definition}

Otherwise stated, a deterministic equivalent for $(f_n)_{n\geq 1}$ is a series $(g_n)_{n\geq 1}$ such that $g_n(z)$ approximates $f_n(\omega,z)$ arbitrarily closely as $n$ grows, for every $z\in D$ and almost every $\omega$. In particular, if $(f_n)_{n\geq 1}$ converges almost surely to a limiting function $f$, i.e., for all $(\omega,z)\in A\times D$ with $A\subset \Omega$, $P(A)=1$ and $D\subset \CC$, we have
\begin{equation}
f_n(\omega,z)\xrightarrow[n\to\infty]{} f(z)
\end{equation}
then $(g_n)_{n\geq 1}$ defined by $g_n=f$, for all $n$, is also a deterministic equivalent of $(f_n)_{n\geq 1}$. In many cases, one can further show that  $\int_{\Omega}f_n(\omega,z)dP(\omega) - g_n(z) \xrightarrow[n\to\infty]{} 0$. Thus $g_n$ is also an approximation of the expected value of $f_n$.

In the context of large dimensional random matrix theory, one often considers random matrices $\Hm_n\in\CC^{N\times n}$ of growing dimensions $N,n\to\infty$, where in general $N/n=c_n$ is such that
\begin{equation}
0<\lim\inf_n c_n\leq \lim\sup_n c_n<\infty.
\end{equation}
This simply states that $c_n$ is bounded so that the ratio $N/n$ of the matrix dimensions is never too close to zero or infinity.
Formally, to be in line with Definition~\ref{def:deteq}, we will define random matrices in the following as series $(\Hm_n)_{n\geq 1}=(\Hm_n(\omega))_{n\geq 1}$ of matrices with growing dimensions which are defined on a probability space $(\Omega,\mathcal F,P)$, where every $\omega\in\Omega$ generates the whole sequence $(\Hm_n(\omega))_{n\geq 1}$ and {\it not} only a single matrix $\Hm_n(\omega)$. 

In wireless communications, one is often interested in the behavior of functionals $f_n(\Hm_n,z)$, where $\Hm_n\in\CC^{N\times n}$ is a matrix describing the input-output relation of a wireless channel. In particular, $f_n(\Hm_n,z)=\frac1N\log\det(\Id_N+z\Hm_n\Hm_n\htp)$, $z\in\RR^+$, is the (normalized) mutual information of the MIMO channel $\Hm_n$ between an $n$-antenna transmitter and an $N$-antenna receiver at signal-to-noise ratio (SNR) $z$. Other quantities of interest are the SINR with linear detectors or precoders and the associated rates. The goal of a large system analysis based on RMT is to provide deterministic approximations of these random quantities, which become arbitrarily tight as the system dimensions grow. Thus, deterministic equivalents provide a \emph{deterministic abstraction of the physical layer}. This is particularly interesting for complex channel models which are intractable by exact analysis. 

Deterministic equivalents for functionals of large dimensional random matrices have been considered for a wide range of communication channel models. For instance, in \cite{HAC07}, a deterministic equivalent for the ergodic mutual information of the Rician fading channel model 
$\Hm_n = \Xm_n + \Am_n$
is provided, where $\Xm_n\in\CC^{N\times n}$ has independent entries with zero mean and a  variance profile $\EE[|(\Xm_n)_{ij}|^2]=\sigma^2_{n,ij}$, and $\Am_n\in\CC^{N\times n}$ is a deterministic matrix. In \cite{DUM10}, the deterministic equivalent of \cite{HAC07} is used to determine an asymptotically tight approximation of the ergodic capacity achieving input covariance matrix for the MIMO Rician fading channel.
Deterministic equivalents were then extended to broader classes of wireless channel models, such as the capacity of the frequency-selective MIMO channel \cite{DUP10}, the MIMO MAC with Kronecker correlation \cite{COU09} and the sum rate capacity of linearly precoded broadcast channels under imperfect channel state information \cite{WAG10}. The application of such techniques is therefore very broad as it can simplify the difficult study of communication channels with a various number of random parameters (random channels, unitary precoders, path loss, etc.). Moreover, deterministic equivalents can be used to compute approximate solutions of otherwise intractable optimization problems \cite{HOY10,WAG10,DUM10}.

All of the works mentioned above consider deterministic equivalents for random matrix models created from sums of independent random matrices. In many cases of practical interest, it is however necessary to consider more complex combinations of matrices, such as products or sums of products. These include for example the multi-hop relay channel (Section~\ref{sec:relay}) as well as the double-scattering channel model \cite{gesbert02} (Section~\ref{sec:double}). 
Another recent example is \cite{couillethoydis11}, which considers random beamforming over fading channels, i.e., both the precoding and the channel matrices are assumed to be random. In this work, the authors derive deterministic equivalents of the mutual information and of the SINR with MMSE detection with respect to the random precoding matrices for quasi-static channels.
Then, a second set of deterministic equivalents is found, treating \emph{both} precoders and channel matrices as random. This technique relies on a fundamental result of probability theory, the Fubini theorem. In this article, we explain this approach in detail and generalize it to  the new notion of {\it iterative deterministic equivalents}.

\subsection{The Fubini theorem}
\begin{theorem}[\cite{BIL08}]
	\label{th:fubini}
Let $(\Omega,\mathcal F,P)$ and $(\Omega',\mathcal F',P')$ be two probability spaces. Denote $(\Omega\times \Omega',\mathcal F\times \mathcal F',Q)$ their product space. Let $f:\Omega\times \Omega'\to\RR$ be $(\mathcal F\times \mathcal F')$-integrable. Then
\begin{align*}
	\int_{\Omega\times \Omega'} f(\omega,\omega')dQ(\omega,\omega') &= \int_{\Omega} \left[\int_{\Omega'} f(\omega,\omega')dP'(\omega')\right]dP(\omega) \\ 
	&= \int_{\Omega'} \left[\int_{\Omega} f(\omega,\omega')dP(\omega)\right]dP'(\omega').
\end{align*}
\end{theorem}\bigskip

In particular, consider a set $A\in\mathcal F\times \mathcal F'$. Then, we have from Theorem \ref{th:fubini} that
\begin{align}
	\label{eq:fubini}
	Q(A) &= \int_{\Omega\times \Omega'} \mathbbm{1}_{A}(\omega,\omega')dQ(\omega,\omega') \nonumber \\
	&= \int_{\Omega'} \left[ \int_{\Omega} \mathbbm{1}_{A}(\omega,\omega') dP(\omega)\right]dP'(\omega').
\end{align}
Equation \eqref{eq:fubini} is the core ingredient for the definition of iterative deterministic equivalents: Let $(\Hm_n(\omega))_{n\geq 1}$ and $(\Hm_n'(\omega'))_{n\geq 1}$ be two series of random matrices generated by the spaces $(\Omega,\mathcal F,P)$ and $(\Omega',\mathcal F',P')$, respectively. As in Theorem \ref{th:fubini}, call $Q$ the product-space measure. Let $f_n( (\Hm_n(\omega),\Hm_n'(\omega')),z)$ be a functional of the matrices $\Hm_n(\omega)$ and $\Hm_n'(\omega')$.
Assume that there is a function $\tilde{g}_n(\Hm_n(\omega),z)$, such that, for each $\omega\in A\subset\Omega$ with $P(A)=1$, there exists a subset $B(\omega)\subset\Omega'$ with $P'(B(\omega))=1$, on which
\begin{align}
 f_n( (\Hm_n(\omega),\Hm_n'(\omega')),z) -\tilde{g}_n(\Hm_n(\omega),z) \to 0.
\end{align}
Although $\tilde{g}_n(\Hm_n(\omega),z)$ is a random function (as it depends on $\omega$), it is independent of $\Hm'_n(\omega')$. Thus, we can see $\tilde{g}_n(\Hm_n(\omega),z)$
as a deterministic equivalent of  $f_n( (\Hm_n(\omega),\Hm_n'(\omega')),z)$ with respect to $(\Hm'_n(\omega'))_{n\ge1}$. 
Now, let us assume that there is a second function $g_n(z)$, such that for $\omega\in C \subset \Omega$ with $P(C)=1$,
\begin{align}
 \tilde{g}_n(\Hm_n(\omega),z) - g_n(z) \to 0.
\end{align}
Call $D = \{(\omega,\omega'): \omega\in A\cap C \ ,\ \omega'\in B(\omega) \}\subset \Omega\times \Omega'$, the space on which $f_n( (\Hm_n,\Hm_n'),z)-g_n(z)\to 0$. Then, from \eqref{eq:fubini}, this space has probability
\begin{align}\nonumber
	Q(D) &= \int_{\Omega} \left[ \int_{\Omega'}  \mathbbm{1}_{D}(\omega,\omega') dP'(\omega')\right]dP(\omega)\\\nonumber
&\overset{(a)}{\ge} \int_{A\cap C} \left[ \int_{B(\omega)} \mathbbm{1}_{D}(\omega,\omega') dP'(\omega')\right]dP(\omega)\\\nonumber
&\overset{(b)}{=} \int_{A\cap C} dP(\omega)\\
&\overset{(c)}{=} 1
\end{align}
where $(a)$ is due to $A\cap B \subset \Omega$ and $B(\omega)\subset \Omega'$, $(b)$ follows since $P'(B(\omega))=1$ for $\omega\in A$ and $(c)$ holds since $P(A\cap C) = P(A)+P(C) - P(A\cup C)=1$. 

To summarize, if a deterministic equivalent $g_n$ exists for a functional $f_n$ of a {\it random} series $(\Hm'_n)_{n\geq 1}$ and a {\it deterministic} series $(\Hm_n)_{n\geq 1}$ of matrices, and if additionally it can be proved that this deterministic equivalent holds true for almost every such $(\Hm_n)_{n\geq 1}$ generated by a space $\Omega$, then the latter is also a deterministic equivalent for the random series $( (\Hm_n,\Hm_n'))_{n\geq 1}$. 

This is the mathematical key idea behind our method to derive iterative deterministic equivalents of functionals $f_n((\Hm_n(\omega),\Hm_n'(\omega')),z)$,  of two (or more) random matrices. First, one considers one of the sequences of random matrices, e.g. $(\Hm_n(\omega))_{n\ge 1}$, to be deterministic and derives a deterministic equivalent with respect to $(\Hm_n'(\omega'))_{n\ge 1}$. In the example above, this was the role of the functional $\tilde{g}_n(\Hm_n(\omega), z)$ which is independent of $\Hm'_n(\omega')$. In a second step, one assumes the matrices $(\Hm_n(\omega))_{n\ge 1}$ to be random and derives an iterative deterministic equivalent $g_n(z)$ of $\tilde{g}_n(\Hm_n(\omega), z)$. Of course, this procedure can be carried out for any finite number of random matrices where in each step the ``randomness'' related to one of the matrices is removed. From the above construction, we will call $(g_n)_{n\geq 1}$ an {\it iterative deterministic equivalent}.

In the next section, we present two specific examples of iterative deterministic equivalents with applications to the capacity of multi-hop MIMO relay and double-scattering channels. From now on, all matrices and vectors should be understood as sequences of matrices and vectors with growing dimensions. For notational convenience, we drop the index $n$, e.g. we write $\Hm$ instead of $(\Hm_n)_{n\ge 1}$.

\section{Applications}
\label{sec:applications}

\subsection{Multi-hop relay channel}\label{sec:relay}
\begin{figure}
\centering
 \includegraphics[width=0.9\textwidth]{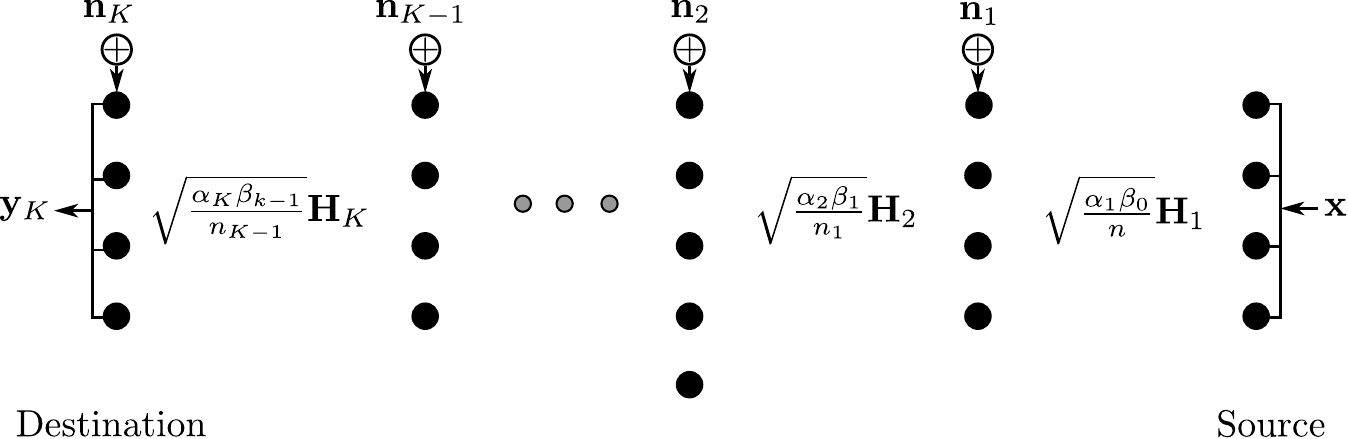}
\caption{Multi-hop amplify-and-forward MIMO relay channel.\label{fig:relaychannel}}
\end{figure}

Consider a multi-hop AF MIMO relay channel where a source node communicates via $K-1$ relays with a destination node. There is no direct link between the source and the destination and each relay can only receive data from the preceding hop. This is for example the case if the nodes follow a time-division multiple-access (TDMA) protocol where only one node is transmitting at any given time and the path loss between relay $k$ and $k-2$ is large. Thus each data symbol reaches the destination after $K$ channel uses. The source and destination are respectively equipped with $n$ and $n_K$ antennas while the $k$th relay has $n_k$ antennas. The relays operate an AF-protocol where each node simply transmits a scaled version of its received signal to the next hop. We will consider a large system limit where $n,n_1,\dots,n_K$ grow infinitely large at the same speed.
Define the following quantities:
\begin{align}\nonumber
 c_1 &= \frac{n}{n_1}\\
 c_k &= \frac{n_{k-1}}{n_k}, \qquad k=2,\dots,K.
\end{align}
The notation ``$n\to\infty$'' must be understood from now on as $n\to\infty$, such that $0<\liminf_n c_k \le \limsup_n c_k < \infty$ for all $k$. We denote $\yv_k\in\CC^{n_k}$ the received base-band signal vector at the $k$th hop, given by
\begin{align}\nonumber
 \yv_1 &= \sqrt{\alpha_1}\Hm_{1} \sqrt{\frac{\beta_{0}}{n}}\xv + \nv_1\\\label{eq:relay}
 \yv_k &= \sqrt{\alpha_k}\Hm_{k} \sqrt{\frac{\beta_{k-1}}{n_{k-1}}}\yv_{k-1} + \nv_k, \qquad k=2,\dots,K
\end{align}
where $\Hm_{k}\in\CC^{n_k\times n_{k-1}}$ is a standard complex Gaussian matrix\footnote{A standard complex Gaussian matrix $\Xm$ has i.i.d.\@ elements $\Xm_{ij}\sim\Cc\Nc(0,1)$.} (let $n_0\defines n$), $\xv\sim\Cc\Nc(\zerov,\Id_n)$ is the channel input vector, $\nv_k\sim\Cc\Nc(\zerov,\Id_{n_k})$ is a noise vector, $\alpha_k$ is a path loss factor, and the parameter $\beta_k$ is chosen to normalize the transmit power of the $k$th node according to its power budget $\rho_k>0$, i.e.,
\begin{align}\nonumber
 \beta_{0}  &= \frac{\rho_0}{\frac1n\trace\mathbb{E}\LSB\xv\xv\htp\RSB} = \rho_0\\\label{eq:beta}
 \beta_{k}  &= \frac{\rho_k}{\frac1{n_k}\trace\mathbb{E}\LSB\yv_{k}\yv_{k}\htp\RSB}, \qquad k=1,\dots,K-1.
\end{align}
The expectation in the last equation is with respect to the transmit and noise vectors only.\footnote{Under a long-term power constraint, the expectation could be taken also with respect to the matrices $\Hm_k$. Asymptotically, both constraints are equivalent (see Lemma~\ref{lem:beta}).} The channel matrices $\Hm_k$ and path loss factors $\alpha_k$ are assumed to be known to the relays and the destination. Since the received signal at each relay is corrupted by noise, the system suffers from noise accumulation. This is in addition to the linear rate loss $\frac1K$ related to the TDMA protocol. Thus, the capacity decreases quickly with the number of hops $K$. Note that our system model is different from existing works which consider either no noise \cite{tucci11}, or noise only at the destination \cite{FAW09}. An exception is \cite{leveque07}, in which the authors consider a similar system model, but do not provide closed-form expressions of the asymptotic mutual information. Several other works deal with the asymptotic capacity of the dual-hop relay channel \cite{morgenshtern06, wagner08}. Recently, an exact expression of the mutual information of the dual-hop channel for finite channel dimensions was derived in \cite{jin10}. Here, we will provide an explicit deterministic equivalent of the mutual information at each relay for the general model \eqref{eq:relay}.

Let us introduce the following, recursively defined matrices $\Rm_{k}\LB\betav_{k-1}\RB$:
\begin{align}\nonumber
\Rm_{0} &= \mathbb{E}\LSB\xv\xv\htp\RSB=\Id_n\\\label{eq:Rk}
 \Rm_{k}\LB\betav_{k-1}\RB &= \mathbb{E}\LSB\yv_{k}\yv_{k}\htp\RSB = \Id_{n_k} + \frac{\alpha_k\beta_{k-1}}{n_{k-1}}\Hm_{k}\Rm_{k-1}\LB\betav_{k-2}\RB\Hm_{k}\htp, \qquad k=1,\dots,K
\end{align}
and the functionals $\Jc_{k}\LB x,\betav_{k-1}\RB$, $x>0$, which are defined as
\begin{align}
\Jc_{k}\LB x,\betav_{k-1}\RB =\frac1{n_k}\log\det\LB\Id_{n_k} + x\frac{\alpha_k\beta_{k-1}}{n_{k-1}}\Hm_{k}\Rm_{k-1}\LB\betav_{k-2}\RB\Hm_{k}\htp \RB,\qquad k=1,\dots,K
\end{align}
where $\betav_{k}=[\beta_{0},\cdots,\beta_{k}]$. 
With these definitions, we can express the normalized mutual information $\Ic_{k}(\betav_{k-1})$ between $\yv_{k}$ and $\xv$ as
\begin{align}\label{eq:mutdiff}
 \Ic_{k}(\betav_{k-1}) = \frac1K\LB\Jc_{k}(1,\betav_{k-1}) - \Jc_{k}(1,\betav'_{k-1})\RB
\end{align}
where $\betav'_{k}=[0,\beta_{1},\cdots,\beta_{k}]$. Next, we demonstrate by a simple example that \eqref{eq:mutdiff} holds.
\bigskip

\begin{example}[2-hop Relay-channel]
The normalized mutual information $\Ic_{2}(\betav_1)$  between $\xv$ and the channel output after the second hop $\yv_{2}$ is given as
\begin{align}\nonumber
 \Ic_{2}(\betav_1) &= \frac1{Kn_2}\log\det\LB\Id_{n_2} + \LB\Id_{n_2} + \frac{\alpha_2 \beta_{1}}{n_1}\Hm_{2}\Hm_{2}\htp\RB^{-1}\frac{\alpha_2\beta_{1}\alpha_1\beta_{0}}{n_1 n}\Hm_{2}\Hm_{1}\Hm_{1}\htp\Hm_{2}\htp\RB\\\nonumber
&= \frac1{Kn_2}\log\det\LB\Id_{n_2} + \frac{\alpha_2 \beta_{1}}{{n_1}}\Hm_{2}\LB\Id_n + \frac{\alpha_1 \beta_{0}}{n}\Hm_{1}\Hm_{1}\htp\RB \Hm_{2}\htp\RB  - \frac1{Kn_2}\log\det\LB\Id_{n_2} + \frac{\alpha_2 \beta_{1}}{n_1}\Hm_{2}\Hm_{2}\htp\RB\\\nonumber
& =  \frac1{Kn_2}\log\det\LB\Id_{n_2} + \frac{\alpha_2 \beta_{1}}{n_1}\Hm_{2}\Rm_{1}(\beta_0) \Hm_{2}\htp\RB - \frac1{Kn_2}\log\det\LB\Id_{n_2} + \frac{\alpha_2 \beta_{1}}{n_1}\Hm_{2}\Rm_{1}(0)\Hm_{2}\htp\RB\\
&=  \frac1K\LB\Jc_{2}(1,\betav_{1}) - \Jc_{2}(1,\betav'_{1})\RB.
\end{align}
\end{example}
\bigskip

In the following, we will derive deterministic equivalents $\bar{\Ic}_{k}(\bar{\betav}_{k-1})$ of $\Ic_{k}(\betav_{k-1})$.
It will turn out that the recursive definition of the matrices $\Rm_{k}(\betav_{k-1})$ in \eqref{eq:Rk} allows us to calculate iterative deterministic equivalents of the mutual information after each hop. This is achieved by treating the matrix $\Rm_{k-1}(\betav_{k-2})$ as deterministic and deriving a deterministic equivalent of $\Jc_K(x,\betav_{k-1})$ with respect to the matrix $\Hm_k$. This process can be iterated for $\Rm_{k-2}(\betav_{k-3}),\Rm_{k-3}(\betav_{k-4}),\dots$ and $\Hm_{k-1},\Hm_{k-2},\dots$ until the deterministic matrix $\Rm_0$ is reached. 
Before we address this problem, we will derive deterministic equivalents $\bar{\beta}_{k}$ of the power normalization factors $\beta_{k}$:

\bigskip
\begin{lemma}[Asymptotic power normalization]\label{lem:beta}
Let $\beta_{0}=\bar{\beta}_0=\rho_0$. Then, 
 \begin{align*}
  \beta_{k} \xrightarrow [n\to\infty]{\text{a.s.}} \bar{\beta}_k =  \frac{\rho_k}{1 + \alpha_k\rho_{k-1}},\qquad k=1,\dots,K-1.
 \end{align*}
\end{lemma}
\bigskip
\begin{IEEEproof}
Recall the definition of $\beta_k = \frac{\rho_k}{\frac{1}{n_k}\trace\Rm_k}$, where $\Rm_k=\Rm_k\LB\betav_{k-1}\RB$. For $k\ge1$, we have
\begin{align}\nonumber
 \frac{1}{n_k}\trace\Rm_k &= 1 + \frac{\alpha_k\beta_{k-1}}{n_kn_{k-1}}\trace\Hm_k\Rm_{k-1}\Hm_k\htp\\\nonumber
& \overset{(a)}{=} 1 + \alpha_k\beta_{k-1}\frac1{n_k}\sum_{j=1}^{n_k}\frac{1}{n_{k-1}}\tilde{\hv}_{k,j}\htp \Rm_{k-1} \tilde{\hv}_{k,j}\\\nonumber
& \overset{(b)}{\asymp} 1 + \alpha_k\frac{\rho_{k-1}}{\frac1{n_{k-1}}\trace\Rm_{k-1}} \frac{1}{n_{k-1}}\trace\Rm_{k-1}\\
& = 1 + \alpha_k \rho_{k-1}
\end{align}
where $(a)$ is obtained by denoting $\tilde{\hv}_{k,j}\in\CC^{n_{k-1}}$ the $j$th row vector of $\Hm_k$ and $(b)$ is due to Lemma~\ref{lem:trace} and Lemma~\ref{lem:norm} in Appendix~\ref{sec:lemmas} and the definition of $\beta_{k-1}$. By the continuous mapping theorem \cite{vdW}, we finally have
\begin{align}
  \beta_{k} = \frac{\rho_k}{\frac{1}{n_k}\trace\Rm_k}  \xrightarrow [n\to\infty]{\text{a.s.}} \frac{\rho_k}{1 + \alpha_k \rho_{k-1}}.
\end{align}
\end{IEEEproof}
\bigskip

In the next theorem, we will build upon Lemma~\ref{lem:beta}, and provide deterministic equivalents of $\Jc_{k}(x,\betav_{k})$.

\bigskip
\begin{theorem}\label{th:mutinfrelay}
For $k\in\{1,\dots,K\}$, let  $\betav_{k-1}=[\beta_{0}\cdots\beta_{k-1}]\ge0$ be a sequence of random vectors, indexed by $n$, and $\bar{\betav}_{k-1}=[\bar{\beta}_0\cdots\bar{\beta}_{k-1}]\ge 0$ be deterministic, such that $\beta_{i}\xrightarrow[n\to\infty]{\text{a.s.}}\bar{\beta}_{i}$ for $i=0,\dots,k-1$. Then,
 \begin{align*}
 \Jc_{k}\LB x,\betav_{k-1}\RB - \bar{\Jc}_k\LB x,\bar{\betav}_{k-1}\RB \xrightarrow [n\to\infty]{\text{a.s.}} 0
\end{align*}
where $ \bar{\Jc}_k\LB x,\bar{\betav}_{k-1}\RB$ is recursively defined for $k\ge 2$ as
\begin{align*}
 \bar{\Jc}_k\LB x,\bar{\betav}_{k-1}\RB  &= c_k\bar{\Jc}_{k-1}\LB\frac{x\alpha_k\bar{\beta}_{k-1}}{c_k+x\alpha_k\bar{\beta}_{k-1}+\bar{e}_{k-1}\LB x,\bar{\betav}_{k-1}\RB},\bar{\betav}_{k-1}\RB + c_k\log\LB1+\frac{x\alpha_k\beta_{k-1}}{c_k + \bar{e}_{k-1}\LB x,\bar{\betav}_{k-1}\RB }\RB\\
 &\qquad + \log\LB1 + \frac{\bar{e}_{k-1}\LB x,\bar{\betav}_{k-1}\RB}{c_k}\RB- \frac{\bar{e}_{k-1}\LB x,\bar{\betav}_{k-1}\RB}{c_k+\bar{e}_{k-1}\LB x,\bar{\betav}_{k-1}\RB}
\end{align*}
and $\bar{e}_{k}\LB x,\bar{\betav}_{k}\RB$ for $k\ge 0$ is given by Theorem~\ref{th:mkek}. The initial value $\bar{\Jc}_1\LB x,\bar{\beta}_{0}\RB$ is given in closed form:
\begin{align*}
 \bar{\Jc}_1\LB x,\bar{\beta}_{0}\RB &= c_1 \log\LB1+\frac{x\alpha_1\bar{\beta}_0}{c_1 + \bar{e}_0\LB x,\bar{\beta}_0\RB}\RB + \log\LB1 + \frac{\bar{e}_0\LB x,\bar{\beta}_0\RB}{c_1}\RB -\frac{\bar{e}_0\LB x,\bar{\beta}_0\RB}{c_1+\bar{e}_0\LB x,\bar{\beta}_0\RB}.
\end{align*}
\end{theorem}\bigskip
\begin{IEEEproof}
 The proof is provided in Appendix~\ref{proof:mutinfrelay}.
\end{IEEEproof}
\bigskip

Theorem~\ref{th:mutinfrelay} allows us to compute the quantities $\bar{\Jc}_k(x,\bar{\betav}_{k-1})$ recursively for any desired relay node $k$. The values of  $\bar{e}_{k-1}\LB x,\bar{\betav}_{k-1}\RB$, needed at each stage, can also be calculated in a recursive manner as shown in the next theorem.

\bigskip
\begin{theorem}\label{th:mkek}
 For $k\in\{1,\dots,K-1\}$, let  $\betav_{k}=[\beta_{0}\cdots\beta_{k}]\ge0$ be a sequence of random vectors, indexed by $n$, and $\bar{\betav}_{k}=[\bar{\beta}_0\cdots\bar{\beta}_{k}]\ge 0$ be deterministic, such that $\beta_{i}\xrightarrow[n\to\infty]{\text{a.s.}}\bar{\beta}_{i}$ for $i=0,\dots,k$. Let $x>0$ and denote by $m_{k}\LB x,{\betav}_{k}\RB=\frac1{n_{k+1}}\trace\LB\alpha_{k+1}\beta_{k}\frac1{n_k}\Hm_{k+1}\Rm_{k}\LB\betav_{k-1}\RB\Hm_{k+1}\htp+\frac1x\Id_{n_{k+1}}\RB^{-1}$. Then,
\begin{align*}
 m_{k}\LB x,\betav_{k}\RB -  \bar{m}_{k}\LB x,\bar{\betav}_k\RB \xrightarrow [n\to\infty]{\text{a.s.}} 0
\end{align*}
where $\bar{m}_{k}\LB x,\bar{\betav}_k\RB$ is recursively defined for $k\ge 1$ as
\begin{align*}
  \bar{m}_{k}\LB x,\bar{\betav}_k\RB = \frac{x c_{k+1}}{c_{k+1}+ \bar{e}_{k}\LB x,\bar{\betav}_k\RB}
\end{align*}
and  $\bar{e}_{k}\LB x,\bar{\betav}_k\RB$ is given as the unique positive solution to the following fixed point equation
\begin{align*}
 \bar{e}_{k}\LB x,\bar{\betav}_k\RB &= c_{k+1}\LB c_{k+1}+\bar{e}_{k}\LB x,\bar{\betav}_k\RB\RB \\&\qquad - \frac{c_{k+1}\LB c_{k+1}+\bar{e}_{k}\LB x,\bar{\betav}_k\RB\RB^2}{x \alpha_{k+1}\bar{\beta}_k}\bar{m}_{k-1}\LB \frac{x\alpha_{k+1} \bar{\beta}_k}{c_{k+1}+x\alpha_{k+1}\bar{\beta}_k + \bar{e}_{k}\LB x,\bar{\betav}_k\RB},\bar{\betav}_{k-1}\RB.
\end{align*}
The initial values $\bar{m}_{0}(x,\bar{\beta}_0)$ and $\bar{e}_{0}(x,\bar{\beta}_0)$ are given in closed form:
\begin{align*}
 \bar{m}_{0}(x,\bar{\beta}_0) & = \frac{c_1}{\frac{\alpha_1\bar{\beta}_0}{c_1+\bar{e}_0\LB x,\bar{\beta}_0\RB}+\frac{1}{x}} +(1-c_1)x\\
\bar{e}_0\LB x,\bar{\beta}_0\RB &= -\frac{ x\alpha_1\bar{\beta}_0(1-c_1) + c_1}{2} + \frac{\sqrt{\LB x\alpha_1\bar{\beta}_0(1-c_1)+c_1\RB^2 + 4x\alpha_1\bar{\beta}_0c_1^2}}{2}.
\end{align*}
\end{theorem}
\bigskip
\begin{IEEEproof}
 The proof is provided in Appendix~\ref{proof:mkek}.
\end{IEEEproof}

\bigskip
\begin{remark}
 The quantity $m_{k}\LB x,{\betav}_{k}\RB$ can be seen as the Stieltjes transform \cite{COUbook} of the empirical spectral distribution (e.s.d.) of the matrix $\alpha_{k+1}\beta_{k}\frac1{n_k}\Hm_{k+1}\Rm_{k}(\betav_{k-1})\Hm_{k+1}\htp$ evaluated at $-\frac1x$. One can further show that Theorem~\ref{th:mkek} implies the weak convergence of the e.s.d. $\alpha_{k+1}\beta_{k}\frac1{n_k}\Hm_{k+1}\Rm_{k}(\betav_{k-1})\Hm_{k+1}\htp$ to a distribution function, whose Stieltjes transform is given by $\bar{m}_{k}$, for almost every $\Hm_1,\dots,\Hm_K$.
\end{remark}
\bigskip

Applying Theorem~\ref{th:mutinfrelay} and Lemma~\ref{lem:beta} to \eqref{eq:mutdiff} yields the following corollary which provides a deterministic equivalent of the mutual information $\Ic_{k}(\betav_{k-1})$:

\bigskip
\begin{cor}[Asymptotic mutual information of the $K$-hop AF MIMO Relay channel]\label{cor:mutinfrelay}
 \begin{align*}
  \Ic_{k}\LB\betav_{k-1}\RB  - \bar{\Ic}_{k}\LB\bar{\betav}_{k-1}\RB \xrightarrow[n\to\infty]{\text{a.s}} 0 \qquad k=1,\dots,K
 \end{align*}
where $$\bar{\Ic}_{k}\LB\bar{\betav}_{k-1}\RB = \frac1K\LB \bar{\Jc}_k(1,\bar{\betav}_{k-1}) - \bar{\Jc}_k(1,\bar{\betav}_{k-1}')\RB$$ with $\bar{\betav}_{k-1}=[\bar{\beta}_0\cdots\bar{\beta}_{k-1}]$, $\bar{\betav}_{k-1}'=[0\ \bar{\beta}_1\cdots\bar{\beta}_{k-1}]$ as given by Lemma~\ref{lem:beta}, and $\bar{\Jc}_k(x,\bar{\betav}_{k-1})$ and  $\bar{\Jc}_k(x,\bar{\betav}'_{k-1})$ as given by Theorem~\ref{th:mutinfrelay}.
\end{cor}
\bigskip

\begin{remark}
 The values of $\bar{\Ic}_{k}(\bar{\betav}_{k-1})$ can be very easily numerically computed. We provide the Matlab code which was used to generate the numerical results in this section in Appendix~\ref{sec:matlab}. Due to the recursive implementation, the computational complexity grows quickly with $k$. Calculating $ \bar{\Jc}_k(x,\bar{\betav}_{k-1})$ with high precision for large values of $k$ ($>10$) seems therefore impractical.
\end{remark}
\bigskip

\begin{figure}
\centering
 \includegraphics[width=0.65\textwidth]{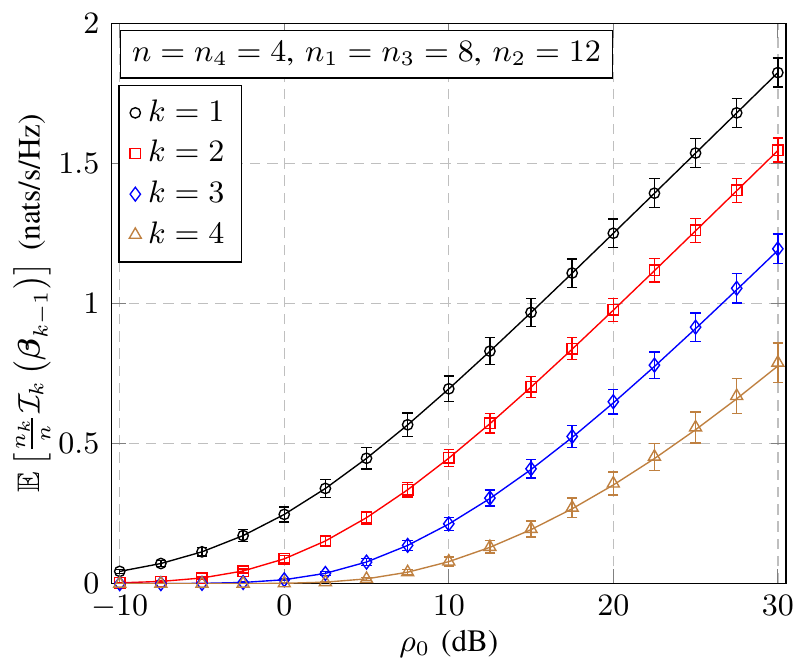}
\caption{Average normalized mutual information $\mathbb{E}\LSB \frac{n_k}{n}\Ic_{k}(\betav_{k-1})\RSB$ after the $k$th hop versus the transmit SNR $\rho_0$ of the source node. The deterministic equivalents $ \frac{n_k}{n}\bar{\Ic}_{k}\LB\bar{\betav}_{k-1}\RB$ are drawn by solid lines, the simulation results by markers. The error bars correspond to one standard deviation of the simulation results in each direction.\label{fig:relay_sim}}
\end{figure}

We would now like to verify our analysis by some numerical results. To this end, we consider a system with three relays, i.e., $K=4$. We assume $n=n_4=4$, $n_1=n_3=8$, $n_2=12$, $\rho_1=\rho_3=0.7\rho_0$ and $\rho_2=0.5\rho_0$. The last assumption allows us to control the transmit power of all nodes by the transmit SNR $\rho_0$ of the source node. We further assume the path loss factors $\alpha_1=1$, $\alpha_2=\alpha_4=0.7$, $\alpha_3=0.5$. Fig.~\ref{fig:relay_sim} shows the average normalized mutual information $\mathbb{E}\LSB \frac{n_k}{n}\Ic_{k}\LB\betav_{k-1}\RB\RSB$ after each hop ($k=1,\dots,4$) versus the transmit power $\rho_0$ of the source node. Note that we have re-normalized all results by $\frac{n_k}{n}$ to put them on a common ground for comparison. The deterministic equivalents $\frac{n_k}{n}\bar{\Ic}_{k}\LB\bar{\betav}_{k-1}\RB$ as provided by Corollary~\ref{cor:mutinfrelay} are drawn by solid lines, simulation results are represented by markers. The error bars represent one standard deviation of the simulation results in each direction. We can observe a very good fit between the asymptotic approximations and the simulation results for all $k$ and the entire range of $\rho_0$. As expected, the performance decreases rapidly with each hop. 

Finally, we would like to remark that, although we have considered a rather simple channel model with neither antenna correlation nor precoding at the nodes, more involved channel models can be treated in a straightforward fashion with the same techniques.

\subsection{Double-scattering MAC}\label{sec:double}
Consider a discrete-time MIMO MAC from $K$ transmitters, equipped with $n_k,\ k=1,\dots,K,$ antennas, respectively, to a receiver with $N$ antennas. The channel output vector $\yv\in\CC^N$ reads
\begin{align}\label{eq:channel}
 \yv &= \sum_{k=1}^K \Hm_{k}\xv_{k} +\nv
\end{align}
where $\Hm_{k}\in\CC^{N\times n_k}$ and $\xv_{k}\in \CC^{n_k}$ are the channel matrix and the transmit vector associated with the $k$th transmitter, $\nv\sim\Cc\Nc(\zerov,\frac1{\rho}\Id_N)$ is a noise vector and $\rho>0$ denotes the SNR. We assume Gaussian signaling, i.e., $\xv_{k}=[x_{k,1},\dots,x_{k,n_k}]\tp\sim\Cc\Nc(\zerov,\Qm_{k})$, where $\Qm_{k}\in\CC^{n_k\times n_k}$. The channel matrices $\Hm_{k}$ are modeled by the double-scattering model \cite{gesbert02}
\begin{align}\label{eqn:model}
 \Hm_{k} = \frac{1}{\sqrt{N_k n_k}}\Rm_{k}^{\frac12}\Wm_{1,k}\Sm_{k}^{\frac12}\Wm_{2,k}\Tm_{k}^{\frac12}
\end{align}
where $\Rm_k\in\CC^{N\times N}$, $\Sm_{k}\in\CC^{N_k\times N_k}$ and $\Tm_k\in\CC^{n_k \times n_k}$ are deterministic correlation matrices, while $\Wm_{1,k}\in\CC^{N\times N_k}$ and $\Wm_{2,k}\in\CC^{N_k\times n_k}$ are independent standard complex Gaussian matrices. Since the distributions of $\Wm_{1,k}$ and $\Wm_{2,k}$ are unitarily invariant, we can assume $\Sm_{k}=\diag(s_{k,1},\dots,s_{k,N_k})$ to be diagonal matrices, without loss of generality for the statistics of $\yv$. 

The double-scattering model \cite{gesbert02} was  motivated by the observation of low-rank channel matrices, despite low antenna correlation at the transmitter and receiver, see e.g.  \cite{muller01,muller02}. A special case of the double-scattering model is the keyhole channel \cite{chizik02,almers06}, which exhibits null antenna correlation, i.e., $\Rm_k=\Id_N$ and $\Tm_k=n_k$ for all $k$, but only a single degree of freedom. The existence of such channels (under laboratory conditions) was confirmed by measurements in \cite{almers06}.
Several theoretical works have studied the double-scattering model so far. The authors of \cite{shin03} derive capacity upper-bounds for the general model and a closed-form expression for the keyhole channel. An asymptotic study of the outage capacity of the multi-keyhole channel was presented in \cite{levin06}.
The diversity order of the double-scattering model was considered in \cite{shin08} and it was shown that a MIMO system with $t$ transmit antennas, $r$ receive antennas and $s$ scatterers achieves the diversity of order $trs/\max(t,r,s)$. A closed-from expression of the diversity-multiplexing trade-off (DMT) was derived in \cite{yang11}. Beamforming along the strongest eigenmode over Rayleigh product MIMO channels, i.e., the double-scattering model without any form of correlation, was considered in \cite{jin08}. Here, the authors derive exact expressions of the cumulative distribution function (cdf) and the probability density function (pdf) of the largest eigenvalue of the Gramian of the channel matrix and compute closed-form results for the ergodic capacity, outage probability and SINR distribution. In a later paper \cite{li10}, the MIMO MAC with double-scattering fading is analyzed. The authors obtain closed-form upper-bounds on the sum-capacity and prove that the transmitters should send their signals along the eigenvectors of the transmit correlation matrices in order to achieve capacity. 
Despite the significant interest in the double-scattering channel model, little work has been done to study its asymptotic performance when the channel dimensions grow large. We are only aware of \cite{muller02}, in which a model without transmit and receive correlation is studied relying on tools from free probability theory. Implicit expressions of the asymptotic mutual information and the SINR with MMSE detection are found therein. 

In the following, we provide deterministic equivalents of the mutual information, the SINR with MMSE-detection and the associated sum-rate. In addition, we derive the precoders which maximize the deterministic equivalent of the  mutual information and provide a simple algorithm for their computation.
The key idea behind the following proofs is that the double-scattering channel can be interpreted as a Kronecker channel \cite{COU09} with a \emph{random} receive correlation matrix, which itself is modeled by the Kronecker model. This observation allows us to build upon \cite{COU09} which provides an asymptotic analysis of the performance of Kronecker channels with deterministic correlation matrices (Theorem~\ref{th:mutinf} in Appendix~\ref{sec:lemmas}). Based on the Fubini theorem, we extend this work by allowing the correlation matrices to be random.
A similar technique can be applied to more involved channel models, such as channels with line-of-sight components or MIMO product channels with an arbitrary number of matrices.

Denote $I_N(\rho)$ the instantaneous normalized mutual information of the channel \eqref{eq:channel}, defined as \cite{COV06}
\begin{align}
 I_N\LB\rho\RB = \frac1N\log\det\LB\Id_N+\rho\sum_{k=1}^K\Hm_k\Qm_k\Hm_k\htp\RB.
\end{align}
Moreover, denote $\gamma^N_{k,j}\LB\rho\RB$ the SINR at the output of the MMSE detector related to the transmit symbol $x_{k,j}$, given by \cite{kaybook}
\begin{align}
 \gamma^N_{k,j}\LB\rho\RB = \hv_{k,j}\htp\LB\sum_{i=1}^K\Hm_i\Hm_i\htp - \hv_{k,j}\hv_{k,j}\htp + \frac1{\rho}\Id_N\RB^{-1}\hv_{k,j}.
\end{align}
We define the normalized sum-rate $R_N\LB\rho\RB$ with MMSE detection as
\begin{align}
 R_N\LB\rho\RB = \frac1N\sum_{k=1}^K\sum_{j=1}^{n_k}\log\LB1+\gamma^N_{k,j}\LB\rho\RB\RB.
\end{align}

The  notation ``$N\to\infty$'' will be used to denote that $N$ and all $N_k$, $n_k$ grow infinitely large, satisfying $ 0 < \lim\inf\frac{N_k}{N} \le \lim\sup \frac{N_k}{N} < \infty$ and $ 0 < \lim\inf\frac{n_k}{N} \le \lim\sup \frac{n_k}{N} < \infty$. Additionally, we need the following technical assumption:

\bigskip\begin{assumption} For all $k$, $\limsup_N\lVert\Rm_k\rVert<\infty$, $\limsup_N\lVert\Sm_k\rVert<\infty$ and $\limsup_N\lVert\Tm_k\Qm_k\rVert<\infty$.
\end{assumption}\bigskip

\begin{remark}
 This assumption implies in particular that the antenna correlation at the transmitter and receiver side cannot grow with the system size, as it would be the case for very dense antenna arrays \cite{GES02}. Amendments to relax this assumption can be made, following the work in \cite{COU09}. Moreover, the last constraint, $\limsup_N\lVert\Tm_k\Qm_k\rVert<\infty$, implies that no transmitter is allowed to focus an increasing amount of transmit power in a single direction.
\end{remark}
\bigskip

Our first theorem introduces a set of $3K$ implicit equations which uniquely determines some quantities $\LB g_k,\bar{g}_k,\delta_k\RB$  ($1\le k\le K$). These will be needed later on to provide deterministic equivalents of $I_N(\rho)$, $\gamma^N_{k,j}\LB\sigma^2\RB$, and $R_N\LB\rho\RB$.

\bigskip
\begin{theorem}[Fundamental equations]\label{th:fundequ}
The following system of $3K$ implicit equations in $\bar{g}_k$, $g_k$, and $\delta_k$ ($1\le k \le K$):
\begin{align}\nonumber
 \bar{g}_k &= \frac{1}{n_k}\trace\Tm_k^{\frac12}\Qm_k\Tm_k^{\frac12}\LB g_k\Tm_k^{\frac12}\Qm_k\Tm_k^{\frac12}+\Id_{n_k}\RB^{-1}\\\label{eq:fundequ}
 g_k &= \frac{1}{n_k}\sum_{j=1}^{N_k}\frac{s_{k,j}\delta_k}{1+\bar{g}_k s_{k,j}\delta_k}\\\nonumber
 \delta_k &= \frac{1}{N_k}\trace\Rm_k\LB\sum_{i=1}^K \frac{n_i}{N_i}\frac{\bar{g}_i g_i}{\delta_i}\Rm_i+\frac1{\rho}\Id_N\RB^{-1}
\end{align}
has a unique solution satisfying $\bar{g}_k,g_k,\delta_k>0$ for all $k$ and $\rho>0$.
\end{theorem}
\begin{IEEEproof}
The proof is provided in Appendix~\ref{proof:fundeq}.
\end{IEEEproof}
\bigskip

\begin{remark}
 The values of $\bar{g}_k$, $g_k$, and $\delta_k$ can be computed by a standard fixed-point algorithm which iteratively computes \eqref{eq:fundequ}, starting from some arbitrary initialization $\bar{g}_k^{(0)},g_k^{(0)},\delta_k^{(0)}>0$. This algorithm is proved to converge, generally terminates within a few iterations (depending on the system size and the desired accuracy), and does not pose any computational challenge.
\end{remark}\vspace{10pt}

The next theorem provides a deterministic equivalent of the (ergodic) mutual information based on the quantities $(g_k,\bar{g}_k,\delta_k)$ as provided by Theorem~\ref{th:fundequ}.

\bigskip\begin{theorem}[Mutual information]\label{th:mutinfscatter}
Assume that $ \textbf{A 1}$ holds. Then,
\begin{align*}
 (i)\quad &I_N\LB\rho\RB-\bar{I}_N\LB\rho\RB \xrightarrow[N\to\infty]{\text{a.s.}} 0\\
 (ii)\quad &\mathbb{E}\LSB I_N\LB\rho\RB\RSB-\bar{I}_N\LB\rho\RB \xrightarrow[N\to\infty]{} 0
\end{align*}
where
\begin{align}\nonumber
  \bar{I}_N\LB\rho\RB  =&\  \frac1N\log\det\LB\Id_N + \rho\sum_{k=1}^K\frac{n_k}{N_k}\frac{\bar{g}_k g_k}{\delta_k}\Rm_k\RB\\\label{eq:mutinfeq}
 &\ + \frac1N\sum_{k=1}^K\LSB \log\det\LB\Id_{N_k}+\bar{g}_k\delta_k\Sm_k\RB + \log\det\LB\Id_{n_k} + g_k\Tm_k^{\frac12}\Qm_k\Tm_k^{\frac12}\RB 
 -2n_k g_k\bar{g}_k\RSB
\end{align}
and $g_k,\bar{g}_k,\delta_k$ are the unique positive solutions to \eqref{eq:fundequ}.
\end{theorem}
\begin{IEEEproof}
The proof is provided in Appendix~\ref{proof:mutinfscatter}.
\end{IEEEproof}
\bigskip

The following result allows us to compute the asymptotically optimal precoding matrices $\Qm_k$ which maximize $\bar{I}_N\LB\rho\RB$ under individual transmit power constraints.

\bigskip\begin{theorem}[Optimal power allocation]\label{th:optpow}
The solution to the following optimization problem:
\begin{align*}
 \LB\bar{\Qm}_1^{{\star}},\dots,\bar{\Qm}_K^{\star}\RB\ =\ & \arg\max_{\Qm_1,\dots,\Qm_k} \bar{I}_N\LB\rho\RB\\
 &\ \text{s.t.}\quad \frac{1}{n_k}\trace \Qm_k\le P_k\ \forall k
\end{align*}
where $\bar{I}_N\LB\rho\RB$ is defined in Theorem~\ref{th:mutinfscatter}, is given as $\bar{\Qm}_k^{\star} = \Um_k\bar{\Pm}_k^{\star} \Um_k\htp$, where $\Um_k\in\CC^{n_k\times n_k}$ is defined by the spectral decomposition of $\Tm_k=\Um_k\diag(t_{k,1},\dots,t_{k,n_k})\Um_k\htp$ and $\bar{\Pm}_k^{\star}=\diag(\bar{p}^{\star}_{k,1},\dots,\bar{p}^{\star}_{k,n_k})$ is given by the water-filling solution:
\begin{align}\label{eq:wf}
 \bar{p}^{{\star}}_{k,j} = \LB\mu_k - \frac{1}{g_k^{\star}t_{k,j}}\RB^+
\end{align}
where $\mu_k$ is chosen to satisfy $\frac1{n_k}\trace\bar{\Pm}_k^{\star}=P_k$ and $g_k^{\star}$ is given by Theorem~\ref{th:fundequ} for $\Qm_k=\bar{\Qm}^{\star}_k$.
\end{theorem}
\begin{IEEEproof}
The proof is provided in Appendix~\ref{proof:optpow}.
\end{IEEEproof}\bigskip

\begin{remark}
 The optimal power allocation matrices $\bar{\Pm}_k^{\star}$  can be calculated by the iterative water-filling Algorithm~1 (see \cite[Remark 2]{COU09} and \cite[Remark 3]{couillethoydis11} for a discussion on the convergence of this algorithm).
\begin{algorithm}
\caption{Iterative water-filling algorithm}
\label{alg:wf}
 \begin{algorithmic}[1]
 \STATE Let $\epsilon>0$, $n=0$ and $\bar{p}^{{\star},0}_{k,j}=P_k$ for all $k,j$.
 \REPEAT
 \STATE For all $k$, compute $g_k^{{\star},n}$ according to Theorem~\ref{th:fundequ} with matrices $\Qm_k=\Um_k\diag\LB \bar{p}_{k,j}^{{\star},n}\RB\Um_k\htp$.
 \STATE For all $k,j$, calculate  $\bar{p}_{k,j}^{{\star},n+1}=\LB\mu_k - \frac{1}{g_k^{{\star},n}t_{k,j}}\RB^{+}$, with $\mu_k$ such that $\frac1{n_k}\sum_{j=1}^{n_k}\bar{p}_{k,j}^{{\star},n+1}=P_k$.
 \STATE $n=n+1$
 \UNTIL{$\max_{k,j} |\bar{p}_{k,j}^{{\star},n}-\bar{p}_{k,j}^{{\star},n-1}|\le\epsilon$} 
\end{algorithmic}
\end{algorithm}
\end{remark}

\begin{remark}
 Denote by $(\Qm_1^{\star},\dots,\Qm_K^{\star})$ the set of precoding matrices which maximize $\mathbb{E}\LSB I_N(\rho)\RSB$ for a given set of power constraints. If the condition $\limsup\lVert\Tm_k\Qm_k^{\star} \rVert<\infty$ holds for all $k$, then $\mathbb{E}\LSB I_N(\rho,\Qm_1^{\star},\dots,\Qm_K^{\star})\RSB - \bar{I}_N(\rho,\bar{\Qm}_1^{{\star}},\dots,\bar{\Qm}_K^{\star})\xrightarrow[N\to\infty]{} 0$, by Theorem~\ref{th:mutinfscatter} and the strict concavity of $\bar{I}_N(\rho)$ and $I_N(\rho)$ in the matrices $\Qm_k$. However, this condition is difficult to verify and is outside the scope of this paper. See \cite{DUM10} for such a technical discussion in the case of Rician fading channels. 
\end{remark}\bigskip

Next, we provide deterministic equivalents of the SINR $\gamma^N_{k,j}\LB\rho\RB $ at the output of the MMSE detector  and the associated sum-rate $R_N\LB\rho\RB$.

\bigskip\begin{theorem}[SINR of the MMSE detector]\label{th:sinr}
Let $\Qm_k=\diag\LB p_{k,1},\dots,p_{k,n_k}\RB$ and $\Tm_k=\diag(t_{k,1},\dots,t_{k,n_k})$ for all $k$. Assume that $\textbf{A 1}$ holds. Then,
\begin{align*}
 \gamma^N_{k,j}\LB\rho\RB-\bar{\gamma}^N_{k,j}\LB\rho\RB\xrightarrow[N\to\infty]{\text{a.s.}} 0
\end{align*}
where $\bar{\gamma}^N_{k,j}\LB\rho\RB= p_{k,j}t_{k,j}g_k $ and $g_k$ is by given by Theorem~\ref{th:fundequ}.
\end{theorem}
\begin{IEEEproof}
 The proof is provided in Appendix~\ref{proof:sinr}.
\end{IEEEproof}\bigskip

\begin{remark}
 It is easy to see that the theorem is also valid under the more general assumption $\Tm_k=\Um_k\diag(t_{k,1},\dots,t_{k,n_k})\Um_k\htp$ and $\Qm_k=\Um_k\diag(p_{k,1},\dots,p_{k,n_k})\Um_k\htp$.
\end{remark}

\bigskip\begin{cor}[Sum-rate with MMSE decoding]\label{cor:rate}
Let $\Qm_k=\diag\LB p_{k,1},\dots,p_{k,n_k}\RB$ and $\Tm_k=\diag(t_{k,1},\dots,t_{k,n_k})$ for all $k$. Assume that $\textbf{A 1}$ holds. Then,
\begin{align*}
 (i)&\quad R_N\LB\rho\RB - \bar{R}_N(\rho) \xrightarrow[N\to\infty]{\text{a.s.}} 0\\
(ii)&\quad \mathbb{E}\LSB R_N\LB\rho\RB\RSB - \bar{R}_N(\rho)\xrightarrow[N\to\infty]{\text{a.s.}} 0
\end{align*}
where 
\begin{align*}
 \bar{R}_N(\rho) = \frac1N\sum_{k=1}^K\sum_{j=1}^{n_k}\log\LB1+\bar{\gamma}^N_{k,j}\LB\rho\RB\RB
\end{align*}
and the $\bar{\gamma}^N_{k,j}\LB\rho\RB$ are given by Theorem~\ref{th:sinr}.
\end{cor}
\begin{IEEEproof}
 The proof is provided in Appendix~\ref{proof:rate}.
\end{IEEEproof}
\bigskip

\begin{remark}
 Careful inspection of \eqref{eq:mutinfeq} reveals that the third term of $\bar{I}_N\LB\rho\RB$ equals $\bar{R}_N(\rho)$ since
\begin{align}
\frac1N\sum_{k=1}^K \log\det\LB\Id_{n_k} + g_k\Tm_k^{\frac12}\Qm_k\Tm_k^{\frac12}\RB =\frac1N\sum_{k=1}^K\sum_{j=1}^{n_k}\log\LB1+p_{k,j}t_{k,j}g_k\RB.
\end{align}
Thus, all other terms in \eqref{eq:mutinfeq} correspond to the gains of successive interference cancellation \cite{tsebook} over simple MMSE detection.
\end{remark}\bigskip

A special case of the double-scattering channel is the Rayleigh product MIMO channel \cite{jin08} which does not exhibit any form of correlation between the transmit and receive antennas or the scatterers. For this model, Theorems~\ref{th:fundequ}, \ref{th:mutinfscatter} and \ref{th:sinr} can be given in closed form as shown in the next corollary.

\bigskip
\begin{cor}[Rayleigh product channel]\label{cor:rayprod}
 For all $k$, let $N_k=S$, $n_k = N$ and assume $\Tm_k=\Id_N$, $\Sm_k=\Id_S$, $\Rm_k=\Id_N$, and $\Qm_k=\Id_N$. Then $\bar{I}_N\LB\rho\RB$ and $ \bar{\gamma}^N_{k,j}\LB\rho\RB$ as defined in Theorems~\ref{th:mutinfscatter} and \ref{th:sinr} can be given in closed form as
\begin{align*}
 \bar{I}_N\LB\rho\RB = \log\LB1+\rho\frac{NK}{S}\bar{g}\LB\bar{g}+\frac{S}{N}- 1\RB\RB- \frac{KS}{N}\log\LB1+\frac NS\LB\bar{g}-1\RB\RB-K\log\LB\bar{g}\RB -2K\LB1-\bar{g}\RB
\end{align*}
and 
\begin{align*}
 \bar{\gamma}^N_{k,j}\LB\rho\RB = \frac{1-\bar{g}}{\bar{g}}
\end{align*}
where $\bar{g}$ is the unique root to
\begin{align}\label{eq:fundequsimple}
 \bar{g}^3 - \bar{g}^2\LB2-\frac SN - \frac1K\RB +\bar{g}\LB1-\frac SN - \frac1K + \frac S{NK}\LB 1+\frac{1}{\rho}\RB\RB - \frac{S}{NK}\frac{1}{\rho} = 0
\end{align}
such that  $\bar{g}\in\LB1-\min\LSB\frac1K,\frac SN\RSB,1\RB$.%$$\delta \defines \frac{1-\bar{g}}{\bar{g}(\bar{g}+S/N-1)}>0$$ and $$g \defines\frac{1-\bar{g}}{\bar{g}}>0.$$
\end{cor}
\begin{IEEEproof}
 The proof is provided in Appendix~\ref{proof:rayprod}.
\end{IEEEproof}
\vspace{10pt}

Note that similar expressions for the asymptotic mutual information and MMSE-SINR have been obtained in \cite{muller02} by means of free probability theory. However, these results require the numerical solution of a third order differential equation.

\begin{figure}
\centering
\includegraphics[width=0.65\textwidth]{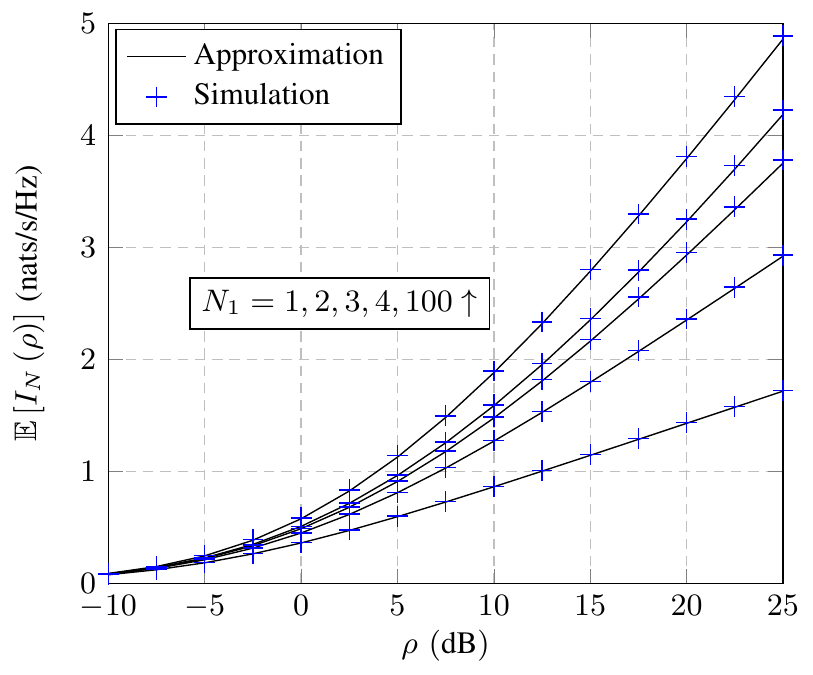}
\caption{Ergodic mutual information $\mathbb{E}\LSB I_N(\rho)\RSB$ of the multi-keyhole channel and its deterministic equivalent $\bar{I}_N(\rho)$ versus $\rho$.\label{fig:keyhole}}
\end{figure}

As a first numerical example, we consider the ``multi-keyhole channel'', i.e., $K=1$, $\Sm_1=\Id_{N_1}$, $\Rm_1=\Id_N$, $\Tm_1=\Qm_1=\Id_{n_1}$,
for $N=n_1=4$. Fig.~\ref{fig:keyhole} depicts the normalized ergodic mutual information $\mathbb{E}\LSB I_N(\rho)\RSB$ and its asymptotic approximation $\bar{I}_N(\rho)$ versus SNR for different numbers of ``keyholes'' $N_1\in\{1,2,3,4,100\}$. Surprisingly, the match between both results is almost perfect although the channel dimensions are very small.
As one expects, the multiplexing gain increases linearly with $N_1$ until $N_1\ge N=4$. Larger values of $N_1$ only change the statistical distribution of the channel matrix  while the degrees of freedom are limited by the number of antennas (for $N_1\to\infty$, $\Hm_1$ becomes a standard Rayleigh fading channel \cite{gesbert02}).

\begin{figure}
\centering
\includegraphics[width=0.65\textwidth]{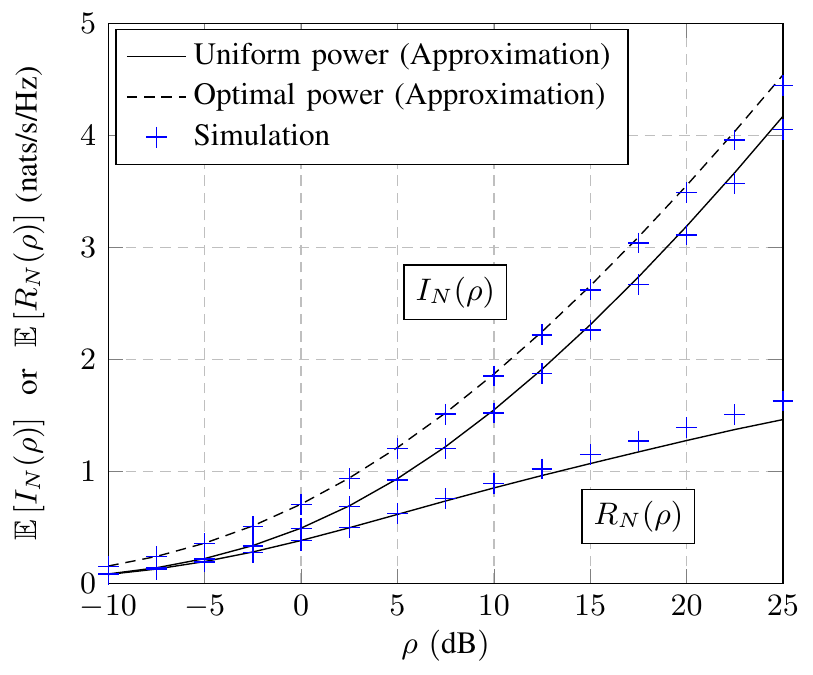}
\caption{Ergodic mutual information $\mathbb{E}\LSB I_N(\rho)\RSB$ and sum-rate $\mathbb{E}\LSB R_N(\rho)\RSB$ of the multiple access channel and their asymptotic approximations $\bar{I}_N(\rho)$ and $\bar{R}_N(\rho)$ versus $\rho$.\label{fig:mac}}
\end{figure}

As a second example, we consider a MAC from $K=3$ transmitters, assuming the double-scattering model in \cite{gesbert02}. Under this model, the correlation matrices are given as  $\Rm_k = \Gm(\phi_{r,k},d_{r,k},N_k)$, $\Sm_k=\Gm(\phi_{s,k},d_{s,k},N_k)$ and $\Tm_k=\Gm(\phi_{t,k},d_{t,k},N_k)$, where $\Gm(\phi,d,n)$ is defined as
\begin{align}
 \LSB\Gm(\phi,d,n)\RSB_{k,l} = \frac1n \sum_{j=\frac{1-n}{2}}^{\frac{n-1}{2}}\exp\LB{\bf{i}}2\pi d(k-l)\sin\LB\frac{j\phi}{1-n}\RB\RB.
\end{align}
The values $\phi_{t,k}$ and  $\phi_{r,k}$ determine the angular spread of the radiated and received signals, $d_{t,k}$ and $d_{r,k}$ are the antenna spacings at the $k$th transmitter and receiver in multiples of the signal wavelength, $N_k$ can be seen as the number of scatterers and $d_{s,k}$ as the spacing of the scatterers. For simplicity, we assume $N=4$, $P_k=1/n_k$, $N_k=11$, $n_k=3$, $d_{t,k}=d_{r,k} = 0.25$ and $d_{s,k}=50$ for all $k$. We further assume $\phi_{r,k}=\phi_{t,k}$ for all $k$, with $\phi_{r,k}\in\{\pi/4,\pi/2,\pi\}$ and $\phi_{s,k}=\pi/8$. Fig.~\ref{fig:mac} shows $\mathbb{E}\LSB I_N(\rho)\RSB$ and $\bar{I}_N(\rho)$ with uniform and optimal power allocation versus SNR. Again, our asymptotic results yield very tight approximations, even for small system dimensions. Note that we have used the  precoding matrices provided by Theorem~\ref{th:optpow} for the simulations as the optimal precoding matrices are unknown. For comparison, we also provide the sum-rate with MMSE detection $\mathbb{E}\LSB R_N(\rho)\RSB$ and its deterministic approximation $\bar{R}_N(\rho)$. We observe a good fit between both results at low SNR values, but a slight mismatch for higher values. This is due to a slower convergence of the SINR $\gamma^N_{k,j}(\rho)$ to its deterministic approximation $\bar{\gamma}^N_{k,j}(\rho)$, well documented in the RMT literature, e.g. \cite{KAM10}.

\section{Conclusion}
\label{sec:conclusion}
In this paper, we have presented a novel tool for the large system analysis of communication systems, called iterative deterministic equivalents. This tool is particularly suited for the analysis of channel models composed of complex combinations of independent random matrices, e.g. products or sums of products of matrices. We have demonstrated the usefulness of this approach with the help of two examples which had not been solved in the literature before. These are the multi-hop AF MIMO relay channel with noise at each stage and the MIMO MAC under the double-scattering channel model. For these channel models, we have provided asymptotically tight deterministic approximations of information theoretic quantities of interest, such as the mutual information and the sum-rate with MMSE detection. These approximations can be easily computed by provably converging fixed-point algorithms and do not require any numerical integration. Our simulation results suggest that the asymptotic performance approximations are very accurate for finite system dimensions with only a few antennas at each node. Finally, the method of iterative deterministic equivalents is applicable to a wide range of channel models of interest (e.g.\@ combinations of correlated i.i.d. and random unitary matrices \cite{couillethoydis11})  which cannot be easily treated so far with other techniques. 

\appendices
\section{Related results}\label{sec:lemmas}
\begin{lemma}[{\cite[Lemma 2.7]{SIL98}, \cite[Lemma 4]{WAG10}}] \label{lem:trace}
 Let $\LB\Am_N\RB_{N\ge 1}$ be a sequence of random $N\times N$ matrices, satisfying $\limsup_N \lVert\Am_N \rVert<\infty$, almost surely. Let $\LB\xv_N\RB_{N\ge 1}$ be a sequence of random $N$-dimensional vectors of i.i.d.\@ entries with zero mean, variance $1/N$ and $8$th order moment of order $O\LB1/N^4\RB$, independent of $\Am_N$. Then,
\begin{align*}
 \xv_N\htp\Am_N\xv_N - \frac1N\trace\Am_N \xrightarrow[N\to\infty]{\text{a.s.}}0.
\end{align*}
\end{lemma}\bigskip

\begin{lemma}[Matrix inversion lemma {\cite[Eq. (2.2)]{SIL95}}]\label{lem:inversion}
 Let $\Am\in\CC^{N\times N}$ be Hermitian invertible. Then, for any vector $\xv\in\CC^N$ and any scalar $\tau\in\CC$ such that $\Am+\tau \xv\xv\htp$ is invertible,
\begin{align*}
 \xv\htp \LB\Am+\tau \xv\xv\htp\RB^{-1} = \frac{\xv\htp\Am^{-1}}{1+\tau \xv\htp \Am^{-1}\xv}. 
\end{align*}
\end{lemma}\bigskip

\begin{lemma}[Rank-$1$ perturbation lemma \cite{SIL95}]\label{lem:rank1perturbation}
Let $z<0$, $\Am\in\CC^{N\times N}$, $\Bm\in\CC^{N\times N}$ with $\Bm$ Hermitian nonnegative definite, and $\vv\in \CC^N$. Then,
\begin{align*}
    \left|\trace\LB(\Bm-z\Id_N)^{-1}-(\Bm+\vv\vv\htp-z\Id_N)^{-1}\RB\Am\right|\leq\frac{\Vert \Am \Vert}{|z|}\ . 
\end{align*}
\end{lemma}\bigskip

\begin{lemma}\label{lem:traceinequ}
 Let $\Rm\in\CC^{N\times N}$ be Hermitian with smallest eigenvalue $\lambda_\text{min}\ge 1$ and $a,b,c,d>0$. Then
\begin{align*}
 \frac1N\trace\Rm\LB a\Rm + b\Id_N\RB^{-1}\Rm\LB c\Rm + d\Id_N\RB^{-1} \ge \frac{1}{(a+b)(c+d)}.
\end{align*}
\begin{IEEEproof}
Let $\Rm = \Um \Deltam \Um\htp$, where $\Um\in\CC^{N\times N}$ is unitary and $\Deltam=\diag\LB\delta_1,\dots,\delta_N\RB \ge 1$. Thus, 
\begin{align}\nonumber
  \frac1N\trace\Rm\LB a\Rm + b\Id_N\RB^{-1}\Rm\LB c\Rm + d\Id_N\RB^{-1} & =  \frac1N\trace\Deltam^2\LB a\Deltam +b\Id_N\RB^{-1}\LB c\Deltam + d\Id_N\RB^{-1}\\\nonumber
&= \frac{1}{(a+b)(c+d)}\frac1N\sum_{i=1}^N \frac{\delta_i^2(a+b)(c+d)}{(a\delta_i + b)(c\delta_i + d)}\\
& \ge \frac{1}{(a+b)(c+d)}.
 \end{align}
\end{IEEEproof}
\end{lemma}

\begin{lemma}\label{lem:norm} Let the matrices $\Rm_{k}(\betav_{k-1}),$ be defined as in \eqref{eq:Rk}. Then, almost surely:
\begin{align*}
 \limsup_{n}\lVert\Rm_{k}(\betav_{k-1})\rVert<\infty,\qquad k=0,\dots,K.
\end{align*}
\end{lemma}
\begin{IEEEproof}
For $k\in\{1,\dots,K\}$, denote by $(\Omega_k,\Fc_k,P_k)$ the probability space generating the sequences of random matrices $\Hm_{k}$.
By \cite{SIL98}, we have on a space $B_k\subset \Omega_k$ with $P_k(B_k)=1$,
\begin{align}
 \frac1{n_{k-1}}\left\lVert\Hm_{k}\Hm_{k}\htp\right\rVert - \LB1+\frac1{\sqrt{c_k}}\RB^2 \xrightarrow [n\to\infty]{}0.
\end{align}
 Obviously, we have $\lVert\Rm_{0}\rVert = \lVert\Id_n \rVert =1$. Thus, almost surely,
\begin{align}
\limsup_n\left \lVert\Rm_{1}(\beta_0)\right\rVert& \le 1+\limsup_n\frac{\alpha_1\beta_0}{n}\left\lVert\Hm_{1}\Hm_{1}\htp\right\rVert = 1+\alpha_1\beta_0 \limsup_n\LB1+\frac1{\sqrt{c_1}}\RB^2< \infty.
\end{align}
Consider now the product space $(\Omega_1\times\Omega_2,\Fc_1\times \Fc_2, Q_2)$. By the Fubini theorem, we have on a subspace $C_2\subset\Omega_1\times\Omega_2 $ with $Q_2(C_2)=1$,
\begin{align}\nonumber
 \limsup_n \lVert\Rm_{2}(\betav_1)\rVert &\le 1+\limsup_n\frac{\alpha_2\beta_{1}}{n_1}\left\lVert\Hm_{2}\Rm_{1}(\beta_0)\Hm_{2}\htp\right\rVert\\\nonumber
& \le  1+ \limsup_n\alpha_2\beta_1\lVert\Rm_{1}(\beta_0)\rVert \frac1{n_1}\left\lVert\Hm_{2}\Hm_{2}\htp\right\rVert\\\nonumber
& = 1 + \alpha_2\beta_1\limsup_n\LB1+\alpha_1\beta_0\LB1+\frac1{\sqrt{c_1}}\RB^2\RB\LB1+\frac1{\sqrt{c_2}}\RB^2\\
&<\infty.
\end{align}
Repeating the last step $k-2$ times concludes the proof.
\end{IEEEproof}\bigskip

\begin{definition}[Standard interference function \cite{yates95}]\label{def:standardfunctions}
A function $\hv(x_1,\ldots,x_K) = [ h_1(x_1,\dots,x_K), \dots,\\h_K(x_1,\ldots,x_K) ]\tp\in \RR^K$ is said to be standard if it fulfills the following conditions:
\begin{enumerate}
	\item {\it Positivity:} for each $j$, if $x_1,\ldots,x_K\geq 0$, then $h_j(x_1,\ldots,x_K)>0$.
	\item {\it Monotonicity:} if $x_1>x_1',\ldots,x_K>x_K'$, then for all $j$, $h_j(x_1,\ldots,x_K)>h_j(x_1',\ldots,x_K')$.
	\item {\it Scalability:} for all $\alpha>1$ and for all $j$, $\alpha h_j(x_1,\ldots,x_K)>h_j(\alpha x_1,\ldots,\alpha x_K)$.
\end{enumerate}
\end{definition}\bigskip

\begin{theorem}[Fixed-point theorem {\cite[Theorem 2]{yates95}}]\label{th:standardfunctions}
If a $K$-variate function $\hv(x_1,\ldots,x_K)$ is standard and there exists $(x_1,\ldots,x_K)$ such that for all $j$, $x_j\geq h_j(x_1,\ldots,x_K)$, then the fixed-point algorithm that consists in setting
\begin{equation*}
	x_j^{(t+1)} = h_j(x_1^{(t)},\ldots,x_K^{(t)})
\end{equation*}
for $t\geq 1$ and for any initial values $x_1^{(0)},\ldots,x_K^{(0)}>0$ converges to the unique jointly positive solution of the system of $K$ equations
\begin{equation*}
x_j = h_j(x_1,\ldots,x_K),\qquad j\in \{1,\ldots,K\}.
\end{equation*}
\end{theorem}\bigskip

\begin{theorem}[{\cite[Corollary 1, Theorem 2]{COU09}}]\label{th:mutinf}
 For $k\in\{1,\dots,K\}$, let $\LB n_k\RB_{N\ge 1}=\LB n_k(N)\RB_{N\ge 1}$ be a sequence of positive integers and let $\LB\Rm_{k,N}\RB_{N\ge1}$, $\Rm_{k,N}\in\CC^{N\times N}$,  $\LB\Tm_{k,N}\RB_{N\ge 1}$,$\Tm_{k,N}\in\CC^{n_k \times n_k}$, and $\LB\Dm_N\RB_{N\ge 1}$, $\Dm_N\in\CC^{N\times N} $, be three sequences of nonnegative definite Hermitian matrices, satisfying $\limsup_N \lVert\Rm_{k,N}\rVert<\infty$, $\limsup_N \lVert\Tm_{k,N}\rVert<\infty$, and $\limsup_N\lVert\Dm_N \rVert<\infty$. Let $\LB\Xm_{k,N}\RB_{N\ge 1}$, $\Xm_{k,N}\in\CC^{N\times n_k}$, be a sequence of random matrices with i.i.d.\@ complex Gaussian entries with zero mean and variance $1/n_k$. Denote $\Bm_N = \sum_k \Rm_{k,N}^{\frac12}\Xm_{k,N}\Tm_{k,N}\Xm_{k,N}\htp\Rm_{k,N}^{\frac12}$
and define the function $V_N(x) = \frac1N\log\det\LB\Id_N + x\Bm_N\RB$ for $x>0$. Let $c_k = n_k/N$ and assume that $0<\liminf_N c_k \le \limsup_N c_k < \infty$ for all $k$. Then,
\begin{align*}
 (i)&\quad\frac1N\trace\Dm_N\LB \Bm_N + \frac{1}{x}\Id_N\RB^{-1} - \frac1N\trace\Dm_N\LB\sum_{i=1}^K \bar{e}_i\Rm_{i,N}+\frac1{x}\Id_N\RB^{-1}\xrightarrow[N\to\infty]{\text{a.s.}}0\\
(ii)&\quad V_N(x) - \bar{V}_N\LB x\RB\xrightarrow[N\to\infty]{\text{a.s.}}0
\end{align*}
where 
\begin{align*}
  \bar{V}_N\LB x\RB &= \frac1N\log\det\LB\Id_N + x \sum_{k=1}^K \bar{e}_{k,N}\Rm_{k,N}\RB + \sum_{k=1}^K \frac1N\log\det\LB\Id_{n_k}+e_{k,N}\Tm_{k,N}\RB -\frac1N\sum_{k=1}^K n_k e_{k,N} \bar{e}_{k,N}
\end{align*}
and where $\bar{e}_{k,N}, e_{k,N},\ k=1,\dots,K$, are given as the unique solution to the equations
\begin{align*}
 \bar{e}_{k,N} &= \frac1{n_k}\trace\Tm_{k,N}\LB e_{k,N}\Tm_{k,N} + \Id_{n_k}\RB^{-1} \\
e_{k,N} &= \frac{1}{n_K} \trace\Rm_{k,N}\LB\sum_{i=1}^K \bar{e}_i\Rm_{i,N}+\frac1{x}\Id_N\RB^{-1}
\end{align*}
such that $\bar{e}_{k,N}, e_{k,N}>0$ for all $k$.
\end{theorem}\bigskip

\begin{cor}[Special case of Theorem~\ref{th:mutinf}, see also \cite{SIL95}]\label{cor:mutinf}
 Let $\LB n\RB_{N\ge 1}=\LB n(N)\RB_{N\ge 1}$ be a sequence of positive integers. Let $\LB\Rm_{N}\RB_{N\ge 1}$, $\Rm_{N}\in\CC^{n \times n}$, be a sequence of nonnegative definite Hermitian matrices, satisfying $\limsup_N \lVert\Rm_{N}\rVert<\infty$ and let $\LB\Xm_{N}\RB_{N\ge 1}$, $\Xm_{N}\in\CC^{N\times n}$, be a sequence of random matrices with i.i.d.\@ complex Gaussian entries with zero mean and variance $1/n$. For $x>0$, define the following functions $m_N\LB x\RB=\frac1N\trace\LB\Xm_N\Rm_N\Xm_N\htp + \frac1x\Id_N\RB^{-1}$ and $J_N\LB x\RB=\frac1N\log\det\LB\Id_N + x\Xm_N\Rm_N\Xm_N\htp\RB$. Denote $c = \frac{n}{N}$ and assume that $0 < \liminf_N c \le \limsup_N c < \infty$. Then,
\begin{align*}
(i)\quad m_N(x) - \bar{m}_N(x)&\xrightarrow [N\to\infty]{\text{a.s.}}0,\qquad (ii)\quad J_N(x) - \bar{J}_N(x)\xrightarrow [N\to\infty]{\text{a.s.}}0
\end{align*}
where
\begin{align*}
\bar{m}_N(x)& = \frac1N\trace\LB\frac{\Rm_N}{c+\bar{e}_N}+\frac{1}{x}\Id_n\RB^{-1} + (1-c)x\\
 \bar{J}_N(x) &= \frac1N\log\det\LB\LSB c+\bar{e}_N\RSB\Id_n + x\Rm_N\RB + (1-c)\log\LB c+\bar{e}_N\RB -\frac{\bar{e}_N}{c+\bar{e}_N} - \log(c)
\end{align*}
and $\bar{e}_N$ is defined as the unique positive solution to the implicit equation
\begin{align}\label{eq:fp}
 \bar{e}_N = \frac1N\trace\Rm_N\LB\frac{\Rm_N}{c+\bar{e}_N}+\frac{1}{x}\Id_n\RB^{-1}.
\end{align}
\end{cor}\bigskip

\section{Proof of Theorem~\ref{th:mutinfrelay}: }\label{proof:mutinfrelay}
First, notice that
\begin{align}\nonumber
\eta_k \defines &\ \left\lVert \Rm_{k}(\betav_{k-1}) - \Rm_{k}(\bar{\betav}_{k-1})  \right \rVert   \\\nonumber
=&\  \left\lVert \frac{\alpha_k \beta_{k-1}}{n_{k-1}}\Hm_k\Rm_{k-1}(\betav_{k-2})\Hm_k\htp - \frac{\alpha_k \bar{\beta}_{k-1}}{n_{k-1}}\Hm_k\Rm_{k-1}(\bar{\betav}_{k-2})\Hm_k\htp \right\rVert\\\nonumber
\le&\ \alpha_k \left\lVert \frac{\Hm_k\Hm_k\htp}{n_{k-1}} \right\rVert \left\lVert \beta_{k-1}\Rm_{k-1}(\betav_{k-2}) - \bar{\beta}_{k-1}\Rm_{k-1}(\bar{\betav}_{k-2})\right\rVert \\\label{eq:diffconv}
\le& \ \alpha_k \left\lVert \frac{\Hm_k\Hm_k\htp}{n_{k-1}}\right\rVert \LSB |\beta_{k-1} - \bar{\beta}_{k-1} |\left\lVert\Rm_{k-1}(\betav_{k-2}) \right\rVert + \bar{\beta}_{k-1} \left\lVert \Rm_{k-1}(\betav_{k-2}) - \Rm_{k-1}(\bar{\betav}_{k-2}) \right\rVert \RSB.
\end{align}
Since, almost surely, $ \limsup\left\lVert \Rm_{1}(\betav_{0}) - \Rm_{1}(\bar{\betav}_{0}) \right\rVert \le \limsup \alpha_1|\beta_0 - \bar{\beta}_0| \left\lVert \frac{\Hm_1\Hm_1\htp}{n}\right\rVert = 0$ and $\limsup\left\lVert\Rm_{k-1}(\betav_{k-2}) \right\rVert<\infty$ (see proof of Lemma~\ref{lem:norm}), one can iteratively show that 
$\eta_k \to 0$, almost surely. Thus,
\begin{align}
\left| \Jc_{k}\LB x,\betav_{k-1}\RB - \Jc_{k}\LB x,\bar{\betav}_{k-1}\RB\right| \xrightarrow [n\to\infty]{\text{a.s.}}&\ 0.
\end{align}
This means that we can from now on replace $\betav_{k}$ by $\bar{\betav}_{k}$ and focus on $\Jc_{k}\LB x,\bar{\betav}_{k-1}\RB$.

As a consequence of Corollary~\ref{cor:mutinf}, Lemma~\ref{lem:norm}, and the Fubini theorem, we obtain the following relation 
\begin{align}
 \Jc_{k}(x,\bar{\betav}_{k-1}) - \tilde{\Jc}_{k}(x,\bar{\betav}_{k-1}) \xrightarrow[n\to\infty]{\text{a.s.}} 0,\qquad k\ge 1
\end{align}
where
\begin{align}\label{eq:recursionJ}\nonumber
 \tilde{\Jc}_{k}\LB x,\bar{\betav}_{k-1}\RB &= \frac1{n_k}\log\det\LB\LSB c_k+e_{k-1}\LB x,\bar{\betav}_{k-1}\RB\RSB\Id_{n_{k-1}}+x\alpha_k\bar{\beta}_{k-1}\Rm_{k-1}\LB\bar{\betav}_{k-2}\RB\RB\\
&\qquad  + (1-c_k)\log\LB c_k + e_{k-1}\LB x,\bar{\betav}_{k-1}\RB\RB - \frac{e_{k-1}\LB x,\bar{\betav}_{k-1}\RB}{c_k+e_{k-1}\LB x,\bar{\betav}_{k-1}\RB} - \log\LB c_k\RB
\end{align}
and $e_{k-1}\LB x,\bar{\betav}_{k-1}\RB$ is given as the unique positive solution to 
\begin{align}
 e_{k-1}\LB x,\bar{\betav}_{k-1}\RB = \frac{1}{n_{k}}\trace\alpha_k\bar{\beta}_{k-1}\Rm_{k-1}\LB\bar{\betav}_{k-2}\RB\LB\frac{\alpha_k\bar{\beta}_{k-1}\Rm_{k-1}\LB\bar{\betav}_{k-2}\RB}{c_k + e_{k-1}\LB x,\bar{\betav}_{k-1}\RB} + \frac1x\Id_{n_{k-1}}\RB^{-1}.
\end{align}

In particular, for $k=1$, we have
\begin{align}\label{eq:resk1}
 \tilde{\Jc}_{1}\LB x,\bar{\beta}_{0}\RB = \bar{\Jc}_1\LB x,\bar{\beta}_{0}\RB
\end{align}
where 
\begin{align}
 \bar{\Jc}_1\LB x,\bar{\beta}_{0}\RB &= c_1\log\LB1+\frac{x\alpha_1\bar{\beta}_0}{c_1+\bar{e}_0\LB x,\bar{\beta}_0\RB}\RB + \log\LB1+ \frac{\bar{e}_0\LB x,\bar{\beta}_0\RB}{c_1}\RB-\frac{\bar{e}_0\LB x,\bar{\beta}_0\RB}{c_1+\bar{e}_0\LB x,\bar{\beta}_0\RB}\\\label{eq:en_closedform}
\bar{e}_0\LB x,\bar{\beta}_0\RB &= -\frac{x\alpha_1\bar{\beta}_0(1-c_1) + c_1}{2} + \frac{\sqrt{\LB x\alpha_1\bar{\beta}_0(1-c_1)+c_1\RB^2 + 4x\alpha_1\bar{\beta}_0c_1^2}}{2}
\end{align}
according to Corollary~\ref{cor:mutinf}. Note that \eqref{eq:fp} permits a closed-form solution \eqref{eq:en_closedform} in this case.

Now, using the recursion $\Rm_{k-1}\LB\bar{\betav}_{k-2}\RB= \Id_{n_{k-1}} + \frac{\alpha_{k-1}\bar{\beta}_{k-2}}{n_{k-2}}\Hm_{k-1}\Rm_{k-2}\LB\bar{\betav}_{k-3}\RB\Hm_{k-1}\htp$ \eqref{eq:Rk} in \eqref{eq:recursionJ}, we obtain
\begin{align}\nonumber
\tilde{\Jc}_{k}(x,\bar{\betav}_{k-1}) &= c_k\Jc_{k-1}\LB\frac{x\alpha_k\bar{\beta}_{k-1}}{c_k+x\alpha_k\bar{\beta}_{k-1}+e_{k-1}\LB x,\bar{\betav}_{k-1}\RB},\bar{\betav}_{k-2}\RB + c_k\log\LB1+\frac{x\alpha_k\bar{\beta}_{k-1}}{c_k +e_{k-1}\LB x,\bar{\betav}_{k-1}\RB}\RB\\\label{eq:111}
&\qquad  + \log\LB1+\frac{e_{k-1}\LB x,\bar{\betav}_{k-1}\RB}{c_k}\RB  - \frac{e_{k-1}\LB x,\bar{\betav}_{k-1}\RB}{c_k+e_{k-1}\LB x,\bar{\betav}_{k-1}\RB}.
\end{align}

In the proof of Theorem~\ref{th:mkek}, it is shown that
\begin{align}
 e_{k-1}\LB x,\bar{\betav}_{k-1}\RB - \bar{e}_{k-1}\LB x,\bar{\betav}_{k-1}\RB \xrightarrow[n\to\infty]{\text{a.s.}}  0.
\end{align}
By the continuous mapping theorem \cite{vdW}, we therefore have
\begin{align}\nonumber
 &\Jc_{k-1}\LB\frac{x\alpha_k\bar{\beta}_{k-1}}{1+x\alpha_k\bar{\beta}_{k-1}+ e_{k-1}\LB x,\bar{\betav}_{k-1}\RB},\bar{\betav}_{k-2}\RB \\ &\qquad\qquad - \Jc_{k-1}\LB\frac{x\alpha_k\bar{\beta}_{k-1}}{1+x\alpha_k\bar{\beta}_{k-1}+\bar{e}_{k-1}\LB x,\bar{\betav}_{k-1}\RB},\bar{\betav}_{k-2}\RB\xrightarrow[n\to\infty]{\text{a.s.}} 0.
\end{align}
Applying the last result together with Corollary~\ref{cor:mutinf}, Lemma~\ref{lem:norm}, the continuous mapping theorem and the Fubini theorem to \eqref{eq:111} concludes the proof for $k=2$ since $\tilde{\Jc}_{1}\LB x,\bar{\beta}_{0}\RB = \bar{\Jc}_1\LB x,\bar{\beta}_{0}\RB$ by \eqref{eq:resk1}. The convergence for $k>2$ is shown by successive iterations of the last steps.

\section{Proof of Theorem~\ref{th:mkek}}\label{proof:mkek}
From standard matrix inequalities and \eqref{eq:diffconv}, it follows that
\begin{align}\label{eq:replace}
\left| m_{k}\LB x,\betav_k\RB - m_{k}\LB x,\bar{\betav}_k\RB\right| \le x^2\alpha_{k+1}\left\lVert \frac{\Hm_{k+1}\Hm_{k+1}\htp}{n_k}\right\rVert \left\lVert\beta_k\Rm_k\LB{\betav}_{k-1}\RB  - \bar{\beta}_k\Rm_k\LB\bar{\betav}_{k-1}\RB\right\rVert
\xrightarrow [n\to\infty]{\text{a.s.}} 0.
\end{align}
Thus, we can replace from now on $\beta_{k}$ by $\bar{\beta}_k$, for almost every $(\Hm_1,\dots,\Hm_K)$.

From Corollary~\ref{cor:mutinf}, Lemma~\ref{lem:norm} and the Fubini theorem, it follows that
\begin{align}\label{eq:conv2}
 m_{k}\LB x,\bar{\betav}_k\RB - \tilde{m}_{k}\LB x,\bar{\betav}_k\RB\xrightarrow[n\to\infty]{\text{a.s.}} 0
\end{align}
where
\begin{align}
 \tilde{m}_{k}(x,\bar{\betav}_k) = \frac1{n_{k+1}}\trace\LB\frac{\alpha_{k+1}\bar{\beta}_k\Rm_{k}\LB\bar{\betav}_{k-1}\RB}{c_{k+1}+e_{k}\LB x,\bar{\betav}_k\RB}+\frac1x\Id_n\RB^{-1} + (1-c_{k+1})x
\end{align}
and $e_{k}\LB x,\bar{\betav}_k\RB$ is given as the unique positive solution to
\begin{align}\label{eq:defeq}
 e_{k}\LB x,\bar{\betav}_k\RB = \frac{1}{n_{k+1}}\trace\alpha_{k+1}\bar{\beta}_k\Rm_{k}\LB\bar{\betav}_{k-1}\RB\LB\frac{\alpha_{k+1}\bar{\beta}_k\Rm_{k}\LB\bar{\betav}_{k-1}\RB}{c_{k+1}+e_{k}\LB x,\bar{\betav}_k\RB}+\frac1x\Id_{n_{k}}\RB^{-1}.
\end{align}
In particular, we have $\tilde{m}_{0}(x,\bar{\betav}_k) = \bar{m}_{0}(x,\bar{\betav}_k)$, where
\begin{align}
 \bar{m}_{0}(x,\bar{\beta}_0) & = \frac{c_1}{\frac{\alpha_1\bar{\beta}_0}{c_1+\bar{e}_0\LB x,\bar{\beta}_0\RB}+\frac{1}{x}} +(1-c_1)x\\
\bar{e}_0\LB x,\bar{\beta}_0\RB &= -\frac{ x\alpha_1\bar{\beta}_0(1-c_1) + c_1}{2} + \frac{\sqrt{\LB x\alpha_1\bar{\beta}_0(1-c_1)+c_1\RB^2 + 4x\alpha_1\bar{\beta}_0c_1^2}}{2}.
\end{align}

Replacing $\Rm_{k}\LB\bar{\betav}_{k-1}\RB$ in \eqref{eq:defeq} by its recursive definition $\Rm_{k}\LB\bar{\betav}_{k-1}\RB= \Id_{n_k} + \frac{\alpha_{k}\bar{\beta}_{k-1}}{n_{k-1}}\Hm_{k}\Rm_{k-1}\LB\bar{\betav}_{k-2}\RB\Hm_{k}\htp$ \eqref{eq:Rk} yields after straightforward calculus
\begin{align}\nonumber
  e_{k}\LB x,\bar{\betav}_k\RB =&\ c_{k+1}\LB c_{k+1} + e_{k}\LB x,\bar{\betav}_k\RB\RB\\&\qquad - \frac{c_{k+1}\LB c_{k+1} + e_{k}\LB x,\bar{\betav}_k\RB\RB^2}{x\alpha_{k+1}\bar{\beta}_k}m_{k-1}\LB\frac{x \alpha_{k+1}\bar{\beta}_k}{c_{k+1} + x \alpha_{k+1}\bar{\beta}_k +e_{k}\LB x,\bar{\betav}_k\RB},\bar{\betav}_{k-1}\RB.
\end{align}

Similarly, one obtains
\begin{align}
 \tilde{m}_{k}(x,\bar{\betav}_k) = \frac{c_{k+1}\LB c_{k+1} + e_{k}\LB x,\bar{\betav}_k\RB \RB}{\alpha_{k+1}\bar{\beta}_k}  m_{k-1}\LB \frac{x\alpha_{k+1}\bar{\beta}_k}{c_{k+1} + x \alpha_{k+1}\bar{\beta}_k + e_{k}\LB x,\bar{\betav}_k\RB  },\bar{\betav}_{k-1}\RB + (1-c_{k+1})x .
\end{align}

Combining the last two equations leads to
\begin{align}
 \tilde{m}_{k}(x,\bar{\betav}_k) = \frac{x c_{k+1}}{c_{k+1}+e_{k}\LB x,\bar{\betav}_k\RB}.
\end{align}

Consider now the quantity $\bar{e}_{k}\LB x,\bar{\betav}_k\RB$, $k\ge1$, defined as a positive solution to 
\begin{align}\nonumber
 \bar{e}_{k}\LB x,\bar{\betav}_k\RB =&\ c_{k+1}\LB c_{k+1}+\bar{e}_{k}\LB x,\bar{\betav}_k\RB\RB\\\label{eq:useonce}&\qquad - \frac{c_{k+1}\LB c_{k+1}+\bar{e}_{k}\LB x,\bar{\betav}_k\RB\RB^2}{x \alpha_{k+1}\bar{\beta}_k}\bar{m}_{k-1}\LB \frac{x\alpha_{k+1} \bar{\beta}_k}{c_{k+1}+x\alpha_{k+1}\bar{\beta}_k + \bar{e}_{k}\LB x,\bar{\betav}_k\RB},\bar{\betav}_{k-1}\RB
\end{align}
where $\bar{m}_{k}\LB x,\bar{\betav}_k\RB$ is recursively defined for $k\ge 1$ as
\begin{align}
  \bar{m}_{k}\LB x,\bar{\betav}_k\RB = \frac{x c_{k+1}}{c_{k+1}+ \bar{e}_{k}\LB x,\bar{\betav}_k\RB}.
\end{align}
It remains to show that a unique solution to \eqref{eq:useonce} exists and that ${e}_{k}\LB x,\bar{\betav}_k\RB-\bar{e}_{k}\LB x,\bar{\betav}_k\RB\xrightarrow[n\to\infty]{\text{a.s.}} 0$. Let us first define the following functions for $k\ge 1$: 
\begin{align}
 f_k(z) &= c_{k+1}\LB c_{k+1} + z\RB - \frac{c_{k+1}\LB c_{k+1} + z\RB^2}{x\alpha_{k+1}\bar{\beta}_k}m_{k-1}\LB\frac{x \alpha_{k+1}\bar{\beta}_k}{c_{k+1} + x \alpha_{k+1}\bar{\beta}_k +z},\bar{\betav}_{k-1}\RB\\
\bar{f}_k(z) &= c_{k+1}\LB c_{k+1} + z\RB - \frac{c_{k+1}\LB c_{k+1} + z\RB^2}{x\alpha_{k+1}\bar{\beta}_k}\bar{m}_{k-1}\LB\frac{x \alpha_{k+1}\bar{\beta}_k}{c_{k+1} + x \alpha_{k+1}\bar{\beta}_k +z},\bar{\betav}_{k-1}\RB.
\end{align}

From \eqref{eq:defeq} and with the help of Lemma~\ref{lem:traceinequ} (note that the smallest eigenvalue of $\Rm_k$ is greater or equal to $1$ for all $k$), one can easily verify that $f_k(z)$ satisfies the following properties for $z\ge0$:
\begin{enumerate}
 \item[(i)] $$f_k(z) \ge c_{k+1}(c_{k+1}+z)\LSB1-\frac{c_{k+1} +z}{c_{k+1} + z + x\alpha_{k+1}\bar{\beta}_k}\RSB > 0$$
\item[(ii)] for $z>z'\ge0$, 
\begin{align*} &\ f_k(z)-f_k(z')\\ 
\ge &\ \frac{(z-z')c_{k+1}\alpha_{k+1}^2\bar{\beta}_k^2}{(c_{k+1}+z')(c_{k+1}+z)}\frac{1}{n_k}\trace\Rm_k\LB\frac{\alpha_{k+1}\bar{\beta}_k\Rm_{k}}{c_{k+1}+z}+\frac1x\Id_{n_{k}}\RB^{-1}\Rm_k\LB\frac{\alpha_{k+1}\bar{\beta}_k\Rm_{k}}{c_{k+1}+z'}+\frac1x\Id_{n_{k}}\RB^{-1}\\ 
\ge&\ \frac{(z-z')c_{k+1}\alpha_{k+1}^2\bar{\beta}_k^2}{\LB \alpha_{k+1}\bar{\beta}_k+ \frac{c_{k+1}+z'}{x}\RB\LB \alpha_{k+1}\bar{\beta}_k + \frac{c_{k+1}+z}{x}\RB}\\
> &\ 0 
\end{align*}
\item[(iii)] for $\alpha>1$, \begin{align*}
               &\ \alpha f_k(z) - f_k(\alpha z) \\
\ge&\ \frac{(\alpha-1)c_{k+1}^2\alpha_{k+1}^2\bar{\beta}_k^2}{(c_{k+1}+\alpha z)(\alpha c_{k+1} + \alpha z)}\frac{1}{n_k}\trace\Rm_k\LB\frac{\alpha_{k+1}\bar{\beta}_k\Rm_{k}}{\alpha c_{k+1}+\alpha z}+\frac1{\alpha x}\Id_{n_{k}}\RB^{-1} \Rm_k\LB\frac{\alpha_{k+1}\bar{\beta}_k\Rm_{k}}{c_{k+1}+\alpha z}+\frac1x\Id_{n_{k}}\RB^{-1}\\
&\quad + \frac{(\alpha-1)c_{k+1}\alpha_{k+1}\bar{\beta}_k}{\alpha x} \frac{1}{n_k}\trace\Rm_k\LB\frac{\alpha_{k+1}\bar{\beta}_k\Rm_{k}}{\alpha c_{k+1}+\alpha z}+\frac1{\alpha x}\Id_{n_{k}}\RB^{-1} \LB\frac{\alpha_{k+1}\bar{\beta}_k\Rm_{k}}{c_{k+1}+\alpha z}+\frac1x\Id_{n_{k}}\RB^{-1}\\
\ge &\ \frac{(\alpha-1)c_{k+1}^2\alpha_{k+1}^2\bar{\beta}_k^2}{\LB\alpha_{k+1}\bar{\beta}_k + \frac{c_{k+1}+\alpha z}{x}\RB\LB\alpha_{k+1}\bar{\beta}_k +\frac{\alpha c_{k+1} + \alpha z}{x}\RB} + \frac{(\alpha-1)c_{k+1}\alpha_{k+1}\bar{\beta}_k}{\alpha x \LB\frac{\alpha_{k+1}\bar{\beta}_k}{\alpha c_{k+1}+\alpha z} + \frac{1}{\alpha x}\RB\LB\frac{\alpha_{k+1}\bar{\beta}_k}{c_{k+1} \alpha z} + \frac1x\RB} \\
 > &\  0
                            \end{align*}
\end{enumerate}
where $\Rm_k = \Rm_k\LB\bar{\betav}_{k-1}\RB$. All properties are independent of $\Rm_k$ and therefore hold for $n\to\infty$.

Assume now $k=1$. For any sequence of bounded non-negative real numbers $z_n$, we have by \eqref{eq:conv2} and the continuous mapping theorem \cite{vdW},
\begin{align}
 f_1(z_n) - \bar{f}_1(z_n) \xrightarrow[n\to\infty]{\text{a.s.}} 0.
\end{align}
Thus, properties $(i)-(iii)$ of $f_1(z)$ also hold for $\bar{f}_1(z)$. By Definition~\ref{def:standardfunctions} and Theorem~\ref{th:standardfunctions}, these properties imply the uniqueness of positive solutions to the fixed point equations $z=f_1(z)$ and $y=\bar{f}_1(y)$, and hence the uniqueness of solutions to \eqref{eq:useonce} for $k=1$.
Moreover, note that
\begin{align}
 |f_k(a) - f_k(b)| \le \frac{\alpha_{k+1}^2\bar{\beta}_k^2 x^2}{c_{k+1}}\lVert \Rm_k\LB\bar{\betav}_{k-1}\RB\rVert^2 |a-b|.
\end{align}
Hence,
\begin{align}\nonumber
 \left| \bar{e}_{1}\LB x,\bar{\betav}_1\RB -  e_{1}\LB x,\bar{\betav}_1\RB \right| &= \left| \bar{f}_1\LB\bar{e}_{1}\LB x,\bar{\betav}_1\RB\RB- f_1\LB e_{1}\LB x,\bar{\betav}_1\RB\RB \right| \\\nonumber
&\le \left| \bar{f}_1\LB\bar{e}_{1}\LB x,\bar{\betav}_1\RB\RB- f_1\LB \bar{e}_{1}\LB x,\bar{\betav}_1\RB\RB \right| + \left| {f}_1\LB\bar{e}_{1}\LB x,\bar{\betav}_1\RB\RB- f_1\LB e_{1}\LB x,\bar{\betav}_1\RB\RB \right|\\
&\le \epsilon_n + \frac{\alpha_{2}^2\bar{\beta}_1^2 x^2}{c_{2}}\lVert \Rm_1\LB\bar{\beta}_{0}\RB\rVert^2 \left| \bar{e}_{1}\LB x,\bar{\betav}_1\RB -  e_{1}\LB x,\bar{\betav}_1\RB \right|
\end{align}
for some sequence of real numbers $\epsilon_n$, satisfying $\epsilon_n\xrightarrow[n\to\infty]{\text{a.s.}} 0.$ By Lemma~\ref{lem:norm}, $\lVert \Rm_1\LB\bar{\beta}_{0}\RB\rVert<M$, almost surely, for some $M>0$. Thus, for $x\le\sqrt{\frac{c_2(1-\delta)}{\alpha_{2}^2\bar{\beta}_1^2 M^2}}$ and some $\delta>0$, we have
\begin{align}
\left| \bar{e}_{1}\LB x,\bar{\betav}_1\RB -  e_{1}\LB x,\bar{\betav}_1\RB \right| \le \frac{\epsilon_n}{\delta} \xrightarrow[n\to\infty]{\text{a.s.}} 0.                                                                                                                                                                                                                                                                                                                               
\end{align}                                                                                                                                                                                                               
Since $\bar{e}_{1}\LB x,\bar{\betav}_1\RB$ and $ e_{1}\LB x,\bar{\betav}_1\RB$ are (almost surely) bounded on any closed subset of $\RR^+\setminus\{0\}$ and have analytic continuations for $x\in\CC\setminus\RR^-$, we have by Vitali's convergence theorem \cite{titchmarsh} that the convergence holds for any $x\in\RR^+\setminus\{0\}$.

The last convergence implies by the continuous mapping theorem that,
\begin{align}
 m_1(x,\bar{\betav}_1) - \bar{m}_{1}\LB x,\bar{\betav}_1\RB \xrightarrow[n\to\infty]{\text{a.s.}} 0.
\end{align}
We now assume $k=2$. The last convergence implies ${f}_{2}(z)\to \bar{f}_{2}(z)$, almost surely. The same steps can therefore be applied to show that $ m_2(x,\bar{\betav}_1) - \bar{m}_{2}\LB x,\bar{\betav}_1\RB \xrightarrow[n\to\infty]{\text{a.s.}} 0.$ This terminates the proof as this process can be iterated $k$ times. 

\section{Proof of Theorem~\ref{th:fundequ}: Fundamental equations}\label{proof:fundeq}
The proof follows essentially the same steps as the proof of Theorem~2 in \cite{couillethoydis11} and will not be given in full detail here. In order to prove the uniqueness of solutions $(\bar{g}_k,g_k,\delta_k)$, it is sufficient to show by Theorem~\ref{th:standardfunctions} that the $K$-variate function $\hv : (x_1,\dots,x_K)\mapsto (h_1,\dots,h_K)$ as defined below, is a standard interference function (see Definition~\ref{def:standardfunctions}).
For $k=1,\dots,K$, we define 
\begin{align}
 h_k(x_1,\dots,x_K) \mapsto \frac{1}{n_k}\sum_{j=1}^{N_k}\frac{s_{k,j}\delta_k}{1+\bar{g}_k s_{k,j}\delta_k}
\end{align}
where \begin{align}\label{eq:needonce}
  \bar{g}_k = \frac{1}{n_k}\trace\Tm_k^{\frac12}\Qm_k\Tm_k^{\frac12}\LB x_k\Tm_k^{\frac12}\Qm_k\Tm_k^{\frac12}+\Id_{n_k}\RB^{-1}     
\end{align}
and $\delta_k,\ k=1,\dots,K,$ form the unique jointly positive solution to the following fixed-point equations 
\begin{align}
\delta_k = \frac{1}{N_k}\trace\Rm_k\LB\sum_{k=1}^K \frac{n_k}{N_k}\frac{\bar{g}_k x_k}{\delta_k}\Rm_k+\frac1\rho\Id_N\RB^{-1}.
\end{align}
The only difference to \cite{couillethoydis11} is the definition of $\bar{g}_k$. In our case, $\bar{g}_k$ is directly defined as a function of $x_k$, whereas $\bar{b}_k$ in  \cite{couillethoydis11} (using their notations) is given as the solution of another fixed point equation. However, the behavior of $\bar{b}_k$ and $\bar{g}_k$ as seen as functions of $x_k$ is identical. In particular, let $x_k > x_k'>0$ and denote by $\bar{g}_k$ and $\bar{g}_k'$ the corresponding values of \eqref{eq:needonce}, respectively. One can easily verify that the following conditions hold:$(i)$ $\bar{g}_k<\bar{g}_k'$ and $(ii)$ $x_k\bar{g}_k > x_k'\bar{g}_k'$. The remaining steps are identical to \cite{couillethoydis11} and will not be repeated here. By showing $\hv(x_1,\dots,x_K)$ to be a standard interference function, we have proved by Theorem~\ref{th:standardfunctions} that the following fixed-point algorithm, which iteratively computes
\begin{align}
 x_k^{t+1} = h_k(x_1^{(t)},\dots,x_K^{(t)}),\qquad k=1,\dots,K
\end{align}
for $t\ge 0$ and some set of initial values $x_1^{(0)},\dots,x_K^{(0)}$, converges as $t\to\infty$ to the unique fixed point $(g_1,\dots,g_K)$.

\section{Proof of Theorem~\ref{th:mutinfscatter}: Mutual information}\label{proof:mutinfscatter}
The key idea is that the double-scattering model can be considered as the Kronecker channel model as considered in Theorem~\ref{th:mutinf} with \emph{random} correlation matrices. Assume now a Kronecker model, for which the matrices $\Hm_k$ are given as
\begin{align}\label{eq:kron}
 \Hm_k = \frac1{\sqrt{n_k}}\Zm_k\Wm_{2,k}\Tm_k^{\frac12}
\end{align}
where $\Zm_k\in\CC^{N\times N_k}$ is a \emph{deterministic} matrix and $\Wm_{2,k}$ and $\Tm_k$ are defined as in \eqref{eqn:model}.
Further assume that $\limsup_N \lVert \Zm_k\rVert< \infty$ for all $k$. Thus, we can apply Theorem~\ref{th:mutinf} to obtain the following deterministic equivalent $\bar{V}_N(\rho)$ of $I_N(\rho)$:
\begin{align}\label{eq:Vn}
 \bar{V}_N(\rho) = \frac1N\log\det\LB\Id_N + \rho\sum_{k=1}^K\bar{e}_k\Zm_k\Zm_k\htp\RB + \sum_{k=1}^K\frac1N\log\det\LB\Id_{n_k}+e_k\Tm_k^{\frac12}\Qm_k\Tm_k^{\frac12}\RB  - \frac1N \sum_{k=1}^K n_ke_k\bar{e}_k
\end{align}
where $\bar{e}_k, e_k, k=1,\dots,K$, are given as the unique solutions to the following equations
\begin{align}\nonumber
 \bar{e}_k &= \frac1{n_k}\trace\Tm_k^{\frac12}\Qm_k\Tm_k^{\frac12}\LB e_k\Tm_k^{\frac12}\Qm_k\Tm_k^{\frac12} + \Id_{n_k}\RB^{-1}\\\label{eq:fundeqkron}
e_k &= \frac1{n_k}\trace\Zm_k\Zm_k\htp\LB\sum_{i=1}^K\bar{e}_i\Zm_i\Zm_i\htp + \frac1{\rho}\Id_N\RB^{-1}
\end{align}
such that $\bar{e}_k, e_k>0$ for all $k$. Recall that the matrices $\Qm_k$ are the covariance matrices of the channel inputs $\xv_k$. Thus, the channel model is equivalent to a channel $\Hm_k = \frac1{\sqrt{n_k}}\Zm_k\Wm_{2,k}\tilde{\Tm}_k^{\frac12}$, where $\tilde{\Tm}_k^{\frac12}=\Tm^{\frac12}\Qm_k^{\frac12}$, with uncorrelated channel inputs $\tilde{\xv}_k\sim\Cc\Nc(0,\Id_{n_k})$.

For the double-scattering channel, the matrices $\Zm_k$ are \emph{random} and defined as
\begin{align}\label{eq:corkron}
 \Zm_k = \frac1{\sqrt{N_k}} \Rm_k^{\frac12}\Wm_{1,k}\Sm_k^{\frac12}.
\end{align}
Let $(\Omega,\Fc,P)$ be the probability space generating the random sequences of matrices $\LB\Wm_{1,k}(\omega)\RB_{N\ge 1}$. There exists $A \subset\Omega$ with $P(A)=1$, such that for each $\omega\in A$, we have $\limsup_N \lVert \Zm_k(\omega)\Zm_k(\omega)\htp \rVert< \infty$ (\cite{SIL98}). Thus, for each of these $\omega$, the matrices $\Zm_k\Zm_k\htp$ satisfy the criteria of the correlation matrices of Theorem~\ref{th:mutinf}. Let $(\Omega',\Fc',P')$ the probability space generating the matrices $\Wm_{2,k}$. Thus, for every $\omega\in A$, there exist a $A'(\omega)\subset \Omega'$ with $P'(A')=1$, such that for all $\omega'\in A'(\omega)$, $\bar{V}_N(\rho)$ is a deterministic equivalent of $I_N(\rho)$. Denote by $(\Omega\times\Omega',\Fc\times\Fc',Q)$ the product space generating the matrices $\Wm_{1,k}(\omega)$ and $\Wm_{1,k}(\omega')$ and denote by $B\subset \Omega\times\Omega'$ the space of all tuples $(\omega,\omega')$, such that $\omega\in A$ and $\omega'\in A'(\omega)$.
By the Fubini theorem, we have $Q(B)=1$, which proves that $\bar{V}_N(\rho)-I_n(\rho)\to 0$, almost surely. However, $\bar{V}_N(\rho)$ is a random quantity, which depends on the matrices $\Zm_k$. Therefore, we will need to obtain an iterative deterministic equivalent $\bar{I}_N(\rho)$ of $\bar{V}_N(\rho)$.

The first step is to replace the fixed-point equations \eqref{eq:fundeqkron} that depend on $\Zm_k$ by deterministic ones. Let us define the quantities $\bar{e}_{k,i,j},\ e_{k,i,j}$, for $ i\in\{1,\dots,K\},\ j\in\{1,\dots,N_k\}$, which are given as the unique solutions to the following set of fixed-point equations:
\begin{align}\nonumber
 \bar{e}_{k,i,j} &= \frac1{n_k}\trace\tilde{\Tm}_k\LB e_{k,i,j}\tilde{\Tm}_k + \Id_{n_k}  \RB^{-1}\\\label{eq:fp_modified}
e_{k,i,j} &= \frac1{n_k}\trace\Zm_{k,i,j}\Zm_{k,i,j}\htp\LB\sum_{\ell=1}^K \bar{e}_{\ell,i,j}\Zm_{\ell,i,j}\Zm_{\ell,i,j}\htp + \frac1{\rho}\Id_N\RB^{-1}
\end{align}
where 
\begin{align*}
 \Zm_{k,i,j} = \begin{cases}
                \Zm_k &, \quad i \ne k\\
		\LSB \zv_{k,1}\cdots \zv_{k,j-1}\zv_{k,j+1}\dots\zv_{k,N_k}\RSB &, \quad i=k
               \end{cases}.
\end{align*}
Obviously, $\bar{e}_{k,i,j}$ and $e_{k,i,j}$ are independent of the vector $\zv_{i,j}$. In addition, we define
\begin{align*}
 Z = \max_k \limsup_N\lVert\Zm_k\Zm_k\htp \rVert,\qquad T = \max_k \limsup_N\lVert\tilde{\Tm}_k \rVert,\qquad n = \min_k n_k, \qquad c = \frac{N}{n}
\end{align*}
and 
\begin{align*}
\alpha_{i,j} = \max_k \left|e_{k,i,j} -e_k \right|,\qquad \bar{\alpha}_{i,j} = \max_k \left|\bar{e}_{k,i,j} -\bar{e}_k \right| .
\end{align*}
Thus, we have for $N$ large,
\begin{align}\nonumber
 \left|\bar{e}_{k,i,j} - \bar{e}_k \right| & = \left|\frac1{n_k}\trace\tilde{\Tm}_k\LB e_{k,i,j}\tilde{\Tm}_k + \Id_{n_k}  \RB^{-1}\LB(e_k - e_{k,i,j})\tilde{\Tm}_{k}\RB\LB e_{k}\tilde{\Tm}_k + \Id_{n_k}  \RB^{-1} \right|\\
&\le\alpha_{i,j} T^2.
\end{align}
Since the right-hand side (RHS) of the last inequality is independent of $k$, we have
\begin{align}\label{eq:inequ3}
 \bar{\alpha}_{i,j} \le \alpha_{i,j} T^2.
\end{align}

On the other hand, for $i\ne k$, we have for $N$ sufficiently large, 
\begin{align}\nonumber
  &\ \left|e_{k,i,j} - e_k \right| \\\nonumber
 = &\ \left|\frac1{n_k}\trace\Zm_k\Zm_k\htp\LB\sum_{\ell=1}^K\bar{e}_{\ell,i,j}\Zm_{\ell,i,j}\Zm_{\ell,i,j}\htp+\frac1{\rho}\Id_N\RB^{-1}\LB\sum_{\ell=1}^K (\bar{e}_\ell-\bar{e}_{\ell,i,j})\Zm_{\ell}\Zm_\ell\htp + \bar{e}_{i,i,j}\zv_{i,j}\zv_{i,j}\htp\RB\LB\sum_{\ell=1}^K\bar{e}_\ell\Zm_\ell\Zm_\ell\htp + \frac1{\rho}\Id_N\RB^{-1} \right|\\\nonumber
\le &\ c K\rho^2 Z^2\bar{\alpha}_{i,j} +  \frac{ \bar{e}_{i,i,j}}{n}\left|\zv_{i,j}\htp\LB\sum_{\ell=1}^K\bar{e}_\ell\Zm_\ell\Zm_\ell\htp + \frac1{\rho}\Id_N\RB^{-1}\Zm_k\Zm_k\htp \LB\sum_{\ell=1}^K\bar{e}_{\ell,i,j}\Zm_{\ell,i,j}\Zm_{\ell,i,j}\htp+\frac1{\rho}\Id_N\RB^{-1}\zv_{i,j}\right|\\\label{eq:inequ1}
\le &\  cK\rho^2 Z^2 \bar{\alpha}_{i,j} + \frac{\rho^2Z^2T}{n}
\end{align}
where the last inequality is due to $\bar{e}_{k,i,j}\le T$ and $\left|\zv_{i,j}\htp \Am \zv_{i,j}\right| \le \lVert\zv_{i,j}\rVert^2 \lVert\Am\rVert \le  Z \lVert\Am \rVert$, for any matrix $\Am$. 

Similarly, one can show that
\begin{align}\label{eq:inequ2}
 \left|e_{k,k,j} - e_k \right| & =  cK\rho^2 Z^2 \bar{\alpha}_{k,j} + \frac{\rho^2Z^2T}{n} + \frac{\rho Z}{n}.
\end{align}
It follows from \eqref{eq:inequ1}, \eqref{eq:inequ2} and \eqref{eq:inequ3}, that
\begin{align}
\alpha_{i,j} \le cK\rho^2 Z^2 \bar{\alpha}_{i,j} + \frac{\rho^2Z^2T}{n} + \frac{\rho Z}{n} \le cK\rho^2 Z^2T^2 \alpha_{i,j} + \frac{\rho^2Z^2T}{n} + \frac{\rho Z}{n} .
\end{align}
Now, for any $\rho \le \sqrt{\frac{1-\epsilon}{cKZ^2T^2}}$ and $\epsilon>0$, we have
\begin{align*}
 \alpha_{i,j} \le \frac{\rho Z}{\epsilon n }\LB1+\rho Z T\RB\qquad , \qquad  \bar{\alpha}_{i,j} \le \frac{\rho Z T^2}{\epsilon n }\LB1+\rho Z T\RB.
\end{align*}
Let $\mu = \max\{\frac{\rho Z}{\epsilon }\LB1+\rho Z T\RB,\frac{\rho Z T^2}{\epsilon }\LB1+\rho Z T\RB\}$, then we finally have
\begin{align}\label{eq:alphas}
 \alpha_{i,j} \le \frac{\mu}{n}\qquad ,\qquad \bar{\alpha}_{i,j}\le \frac{\mu}{n}.
\end{align}

The last result establishes that for sufficiently small $\rho$, the differences between the solutions $(\bar{e}_{k,i,j},e_{k,i,j})$ to \eqref{eq:fp_modified} and the solutions $(\bar{e}_k,e_k)$ to \eqref{eq:fundeqkron} are uniformly bounded by $\mu$ and vanish as $n\to\infty$. Moreover, $e_k$ and $e_{k,i,j}$ have an analytic continuation on  $z=-\frac1{\rho}\in \CC\setminus \RR^+$ and are uniformly bounded on all closed subsets of $z\in\CC\setminus\RR^+$. Thus, in particular for $\rho\in\RR^+\setminus\{0\}$ and all $k,i,j$, we have by the Vitali convergence theorem \cite{titchmarsh}
\begin{align}\label{eq:final_conv}
  e_k - e_{k,i,j} \xrightarrow[N\to\infty]{} 0\qquad \text{and hence}\qquad \bar{e}_k - \bar{e}_{k,i,j}\xrightarrow[N\to\infty]{} 0.
\end{align}

As a consequence of \eqref{eq:final_conv}, we can now write
\begin{align}\nonumber
 e_k &= \frac1{n_k}\sum_{j=1}^{N_k}\zv_{k,j}\htp\LB\sum_{i=1}^K\bar{e}_i\Zm_i\Zm_i\htp + \frac1{\rho}\Id_N\RB^{-1}\zv_{k,j}\\\nonumber
&\overset{(a)}{\asymp} \frac1{n_k}\sum_{j=1}^{N_k}\zv_{k,j}\htp\LB\sum_{i=1}^K\bar{e}_{i,k,j}\Zm_i\Zm_i\htp + \frac1{\rho}\Id_N\RB^{-1}\zv_{k,j}\\\nonumber
& \overset{(b)}{=} \frac1{n_k}\sum_{j=1}^{N_k}\frac{\zv_{k,j}\htp\LB\sum_{i=1}^K\bar{e}_{i,k,j}\Zm_i\Zm_i\htp - \bar{e}_{k,k,j}\zv_{k,j}\zv_{k,j}\htp+ \frac1{\rho}\Id_N\RB^{-1}\zv_{k,j}}{1+\bar{e}_{k,k,j}\zv_{k,j}\htp\LB\sum_{i=1}^K\bar{e}_{i,j,k}\Zm_i\Zm_i\htp - \bar{e}_{k,k,j}\zv_{k,j}\zv_{k,j}\htp+ \frac1{\rho}\Id_N\RB^{-1}\zv_{k,j}}\\\nonumber
&\overset{(c)}{\asymp}  \frac1{n_k}\sum_{j=1}^{N_k} \frac{\frac{s_{k,j}}{N_k}\trace\Rm_k\LB\sum_{i=1}^K\bar{e}_{i,j,k}\Zm_i\Zm_i\htp + \frac1{\rho}\Id_N\RB^{-1}}{1+\frac{s_{k,j}\bar{e}_{k,k,j}}{N_k}\trace\Rm_k\LB\sum_{i=1}^K\bar{e}_{i,j,k}\Zm_i\Zm_i\htp + \frac1{\rho}\Id_N\RB^{-1}}\\
&\overset{(d)}{\asymp}  \frac1{n_k}\sum_{j=1}^{N_k} \frac{\frac{s_{k,j}}{N_k}\trace\Rm_k\LB\sum_{i=1}^K\bar{e}_{i}\Zm_i\Zm_i\htp + \frac1{\rho}\Id_N\RB^{-1}}{1+\frac{s_{k,j}\bar{e}_k}{N_k}\trace\Rm_k\LB\sum_{i=1}^K\bar{e}_{i}\Zm_i\Zm_i\htp + \frac1{\rho}\Id_N\RB^{-1}}
\end{align}
where $(a)$ follows from \eqref{eq:final_conv} since 
\begin{align}\label{eq:replace2}
 \left|\zv_{k,j}\htp\LB\sum_{i=1}^K\bar{e}_i\Zm_i\Zm_i\htp + \frac1{\rho}\Id_N\RB^{-1}\zv_{k,j} - \zv_{k,j}\htp\LB\sum_{i=1}^K\bar{e}_{i,k,j}\Zm_i\Zm_i\htp + \frac1{\rho}\Id_N\RB^{-1}\zv_{k,j} \right|\le \frac{\mu KZ\rho^2}{n}\xrightarrow[N\to\infty]{}0,
\end{align}
$(b)$ is due to Lemma~\ref{lem:inversion}, $(c)$ is a consequence of Lemmas~\ref{lem:trace} and \ref{lem:rank1perturbation} and $(d)$ is obtained by applying \eqref{eq:final_conv} a second time. Next, we would like to find deterministic equivalents of the terms $\frac1{N_k}\trace\Rm_k\LB\sum_{i=1}^K\bar{e}_{i}\Zm_i\Zm_i\htp + \frac1{\rho}\Id_N\RB^{-1}$. 
We cannot directly apply Theorem~\ref{th:mutinf} at this point since the $\bar{e}_{k}$ are defined as functions of $\Zm_k$. However, based on the relations \eqref{eq:alphas} and \eqref{eq:replace2}, Theorem~\ref{th:mutinf} (see \cite[Theorem 1]{COU09}) can be shown to hold also for the matrix model under study.
Thus, 
\begin{align}
 \frac1{N_k}\trace\Rm_k\LB\sum_{i=1}^K\bar{e}_{i}\Zm_i\Zm_i\htp + \frac1{\rho}\Id_N\RB^{-1} \asymp f_k
\end{align}
where $f_k$ for $k\in\{1,\dots,K\}$ are defined as the unique solution to the following fixed-point equations
\begin{align}\label{eq:f1}
 \bar{f}_k &= \frac1{N_k}\sum_{j=1}^{N_k} \frac{s_{k,j}\bar{e}_k}{1+ \bar{e}_k s_{k,j}f_k}\\\label{eq:f2}
f_k &= \frac{1}{N_k}\trace\Rm_k \LB\sum_{i=1}^K \bar{f}_i\Rm_i + \frac1{\rho}\Id_N \RB^{-1}
\end{align}
such that $f_k>0$ for all $k$. Replacing \eqref{eq:f1} in \eqref{eq:f2} leads to
\begin{align}
 f_k &= \frac{1}{N_k}\trace\Rm_k\LB\sum_{i=1}^K \frac{\bar{e}_i}{N_i} \Rm_i\sum_{j=1}^{N_i} \frac{s_{i,j}}{1+s_{i,j}\bar{e}_if_{i}} + \frac{1}{\rho}\Id_N\RB^{-1}.
\end{align}
Thus,
\begin{align}
 e_k &= \frac{1}{n_k}\sum_{j=1}^{N_k} \frac{s_{k,j} f_k}{1+s_{k,j}\bar{e}_k f_k} + \epsilon_k
\end{align}
where $\epsilon_k$ is a sequence of random variables, satisfying $\epsilon_k \xrightarrow[N\to\infty]{\text{a.s.}}0$.
Consider now the following system of equations
\begin{align*}
  \bar{e}_k &= \frac1{n_k}\trace\tilde{\Tm}_k\LB e_k\tilde{\Tm}_k + \Id_{n_k}\RB^{-1} \\
e_k &= \frac{1}{n_k}\sum_{j=1}^{N_k} \frac{s_{k,j} f_k}{1+s_{k,j}\bar{e}_k f_k} + \epsilon_k\\
f_k &=  \frac{1}{N_k}\trace\Rm_k\LB\sum_{i=1}^K \frac{\bar{e}_i}{N_i} \Rm_i\sum_{j=1}^{N_i} \frac{s_{i,j}}{1+s_{i,j}\bar{e}_if_{i}} + \frac{1}{\rho}\Id_N\RB^{-1}
\end{align*}
and its deterministic counterpart
\begin{align*}
   \bar{g}_k &= \frac1{n_k}\trace\tilde{\Tm}_k\LB g_k\tilde{\Tm}_k + \Id_{n_k}\RB^{-1} \\
g_k &= \frac{1}{n_k}\sum_{j=1}^{N_k} \frac{s_{k,j} \delta_k}{1+s_{k,j}\bar{g}_k \delta_k}\\
\delta_k &=  \frac{1}{N_k}\trace\Rm_k\LB\sum_{i=1}^K \frac{\bar{g}_i}{N_i} \Rm_i\sum_{j=1}^{N_i} \frac{s_{i,j}}{1+s_{i,j}\bar{g}_i\delta_{i}} + \frac{1}{\rho}\Id_N\RB^{-1} = \frac{1}{N_k}\trace\Rm_k\LB \sum_{i=1}^K\frac{n_i}{N_i}\frac{\bar{g}_ig_i}{\delta_i}\Rm_i + \frac1{\rho}\Id_N \RB^{-1}.
\end{align*}
Define the quantities:
\begin{align*}
 \gamma_1 = \max_k\left| e_k - g_k \right|,\qquad \gamma_2 = \max_k \left| \bar{e}_k - \bar{g_k} \right|, \qquad \gamma_3 = \max_k\left| f_k - \delta_k \right|,\qquad \epsilon = \max_k \left| \epsilon_k\right|.
\end{align*}
Straight-forward calculations lead to the following bounds:
\begin{align*}
 \gamma_1 \le c S \gamma_3 + cS^2R^2\rho^2 \gamma_2 + \epsilon, \qquad \gamma_2 \le \gamma_1 T^2,\qquad \gamma_3 \le KSR^2\rho^2 \gamma_2 + KS^2R^2T^2\rho^2 \gamma_1.
\end{align*}
 Combining these results and using the fact that $\epsilon\xrightarrow[N\to\infty]{\text{a.s.}}0$ yields for $\rho$ sufficiently small, 
\begin{align}\label{eq:convek}
 \gamma_1 , \gamma_2, \gamma_3 \xrightarrow[N\to\infty]{\text{a.s.}} 0 .
\end{align}
Since $e_k,g_k,\bar{e}_k,\bar{g}_k,f_k,\delta_k$ are all (almost surely) bounded for $\rho$ in any closed subset of $\RR^+\setminus\{0\}$ and have analytic continuations for $\rho\in\CC\setminus \RR^-$, we have by Vitali's convergence theorem \cite{titchmarsh} that \eqref{eq:convek} holds for any $\rho\in\RR^+\setminus\{0\}$.

Coming now back to $\bar{V}_N(\rho)$ as given in \eqref{eq:Vn}, we have from the continuous mapping theorem \cite{vdW}  that
\begin{align}\label{eq:mutconv1}
 \frac1N\sum_{k=1}^K \LSB \log\det\LB\Id_{n_k} +  e_k\tilde{\Tm}_k\RB - n_k e_k \bar{e}_k\RSB -  \LSB\log\det\LB\Id_{n_k} + g_k\tilde{\Tm}_k\RB -n_kg_k\bar{g}_k\RSB \xrightarrow[N\to\infty]{\text{a.s.}}0.
\end{align}
Moreover, since $\lVert\sum_{k=1}^K\LB\bar{e}_k-\bar{g}_k\RB\Zm_k\Zm_k\htp\rVert\xrightarrow[N\to\infty]{\text{a.s.}}0$, we have
\begin{align}
 \frac{1}{N}\log\det\LB\Id_N + \rho\sum_{k=1}^K\bar{e}_k\Zm_k\Zm_k\htp\RB - \frac{1}{N}\log\det\LB\Id_N + \rho\sum_{k=1}^K\bar{g}_k\Zm_k\Zm_k\htp\RB\xrightarrow[N\to\infty]{\text{a.s.}}0.
\end{align}
Applying Theorem~\ref{th:mutinf} to the last term yields
\begin{align}\nonumber
 &\frac{1}{N}\log\det\LB\Id_N + \rho\sum_{k=1}^K\bar{g}_k\Zm_k\Zm_k\htp\RB \\ \label{eq:mutconv2}
&\quad-\frac{1}{N}\log\det\LB\Id_N +\rho\sum_{k=1}^K\frac{n_k}{N_k}\frac{\bar{g}_k g_k}{\delta_k}\Rm_k\RB - \frac{1}{N}\sum_{k=1}^K \log\det\LB\Id_{N_k}+ \bar{g}_k\delta_k \Sm_k\RB + n_k \bar{g}_k g_k\xrightarrow[N\to\infty]{\text{a.s.}}0.
\end{align}
Combining \eqref{eq:mutconv1} and \eqref{eq:mutconv2} finally leads to
\begin{align}
 \bar{V}_N(\rho) - \bar{I}_N(\rho)\xrightarrow[N\to\infty]{\text{a.s.}}0
\end{align}
where
\begin{align}\nonumber
 \bar{I}_N(\rho) &= \frac{1}{N}\log\det\LB\Id_N +\rho\sum_{k=1}^K\frac{n_k}{N_k}\frac{\bar{g}_k g_k}{\delta_k}\Rm_k\RB + \frac{1}{N}\sum_{k=1}^K \log\det\LB\Id_{N_k}+ \bar{g}_k\delta_k \Sm_k\RB - n_k \bar{g}_k g_k\\
&\qquad + \frac{1}{N}\sum_{k=1}^K\log\det\LB\Id_{n_k} + g_k\tilde{\Tm}_k\RB -n_kg_k\bar{g}_k.
\end{align}

This concludes the proof of part $(i)$.

In order to show the convergence in the mean (part $(ii)$), we will pursue the same approach as in \cite{HAC07,couillethoydis11}. Define the following functions:
\begin{align*}
 m_N(z) = \frac1N\trace\LB \sum_{k=1}^K\Hm_k\Hm_k\htp - z\Id_N\RB^{-1},\qquad \bar{m}_N(z) = \frac1N\trace\LB\sum_{k=1}^K \frac{n_k}{N_k}\frac{\bar{g}_kg_k}{\delta_k}\Rm_k -z\Id_N\RB^{-1}.
\end{align*}
One can show that
\begin{align}
 \mathbb{E}\LSB I_N(\rho)\RSB - \bar{I}_N(\rho) = \int_{\frac1{\rho}}^\infty \LB\LSB\frac1\omega - \mathbb{E}\LSB m_N(-\omega)\RSB\RSB - \LSB \frac1\omega - \bar{m}_N(-\omega)\RSB\RB d\omega.
 \end{align}
Since both $m_N(\omega)$ and $\bar{m}_N(\omega)$ are uniformly bounded by $\frac1\omega$, it follows from dominated convergence arguments, Theorem~\ref{th:mutinf} and \eqref{eq:convek} that, for all $\omega>0$,
\begin{align}
 \LSB\frac1\omega - \mathbb{E}\LSB m_N(-\omega)\RSB\RSB - \LSB \frac1\omega - \bar{m}_N(-\omega)\RSB\to 0.
\end{align}
Moreover,
\begin{align}\nonumber
 \left| \LSB\frac1\omega - \mathbb{E}\LSB m_N(-\omega)\RSB\RSB - \LSB \frac1\omega - \bar{m}_N(-\omega)\RSB\right|& \le  \left| \LSB\frac1\omega - \mathbb{E}\LSB m_N(-\omega)\RSB\RSB\right| + \left|\LSB \frac1\omega - \bar{m}_N(-\omega)\RSB\right|\\\nonumber
& \le \frac1{\omega^2}\LB \frac1N\trace\mathbb{E}\LSB \sum_{k=1}^K\Hm_k\Hm_k\htp\RSB +\frac1N\trace\LB\sum_{k=1}^K \frac{n_k}{N_k}\frac{\bar{g}_kg_k}{\delta_k}\Rm_k \RB\RB\\
&\le \frac{2K R S T}{\omega^2} 
\end{align}
where $R = \max_k\limsup\lVert\Rm_k \rVert$, $S = \max_k\limsup\lVert\Sm_k \rVert$, $T = \max_k\limsup\lVert\Tm_k\Qm_k \rVert$. Since $\frac{2K R S T}{\omega^2}$ is finite and integrable over $[\frac1\rho,\infty)$, it follows from the dominated convergence theorem that
\begin{align}
 \mathbb{E}\LSB I_N(\rho)\RSB - \bar{I}_N(\rho) \xrightarrow[N\to\infty]{} 0.
\end{align}

\section{Proof of Theorem~\ref{th:optpow}: Optimal power allocation}\label{proof:optpow}
The proof follows closely those of \cite[Proposition 5]{DUM10} and \cite[Proposition 3]{COU09}. 

We first recall the following property of concave functions (see e.g. \cite{boydcvx}):

\bigskip\begin{property}
 A function $f\LB\Qm_1,\dots,\Qm_K\RB$ is strictly concave in the Hermitian nonnegative matrices $\Qm_1,\dots,\Qm_K$, if and only if, for any couples $\LB\Qm_{1_a},\Qm_{1_b}\RB,\dots,\LB\Qm_{K_a},\Qm_{K_b}\RB$ of Hermitian nonnegative matrices, the function
$$ \phi(\lambda) = f\LB \lambda\Qm_{1_a} + (1-\lambda)\Qm_{1b},\dots,\lambda\Qm_{K_a} + (1-\lambda)\Qm_{Kb}\RB,\qquad \lambda\in[0,1]$$
is strictly concave.
\end{property}\bigskip

Consider now $\bar{I}_N(\rho)$ seen as a function of $\lambda$ for $\Qm_k = \lambda\Qm_{k_a}-(1-\lambda)\Qm_{k_b}$, where $\Qm_{k_a},\Qm_{k_b}$ are Hermitian nonnegative definite matrices, for $k=1,\dots,K$. Thus, by the chain rule of differentiation,
\begin{align}
 \frac{d \bar{I}_N(\rho)}{d \lambda} =  \frac{\partial \bar{I}_N(\rho)}{\partial \lambda} + \sum_{k=1}^K \frac{\partial \bar{I}_N(\rho)}{\partial \bar{g}_k} \frac{\partial \bar{g}_k}{\partial \lambda} + \frac{\partial \bar{I}_N(\rho)}{\partial g_k}\frac{\partial g_k}{\partial \lambda} + \frac{\partial \bar{I}_N(\rho)}{\partial \delta_k}\frac{\partial \delta}{\partial \lambda}.
\end{align}
One can verify that the partial derivatives of $\bar{I}_N\LB\rho\RB$ with respect to $g_k,\bar{g}_k,\delta_k$, respectively, satisfy
\begin{align}\label{eq:partderiv}
 \frac{\partial\bar{I}_N\LB\rho\RB}{\partial g_k} = \frac{\partial\bar{I}_N\LB\rho\RB}{\partial \bar{g}_k} = \frac{\partial\bar{I}_N\LB\rho\RB}{\partial \delta_k} = 0,
\end{align}
due to the defining relation \eqref{eq:fundequ}. Thus,
\begin{align}
  \frac{d \bar{I}_N(\rho)}{d \lambda} =  \frac{\partial \bar{I}_N(\rho)}{\partial \lambda} = \sum_{k=1}^K\frac1N\trace\Tm_{k}^{\frac12}\LB\Id_{n_k} + g_k\Tm_k^{\frac12}\LB\lambda\Qm_{k_a}+(1-\lambda)\Qm_{k_b}\RB\Tm_{k}^{\frac12} \RB^{-1}\Tm_k^{\frac12}\LB\Qm_{k_a}-\Qm_{k_b}\RB.
\end{align}
The second derivative therefore reads
\begin{align}
 \frac{d^2 \bar{I}_N(\rho)}{d \lambda^2} = -\sum_{k=1}^K\frac1N\trace\LSB\underbrace{\Tm_{k}^{\frac12}\LB\Id_{n_k} + g_k\Tm_k^{\frac12}\LB\lambda\Qm_{k_a}+(1-\lambda)\Qm_{k_b}\RB\Tm_{k}^{\frac12} \RB^{-1}\Tm_k^{\frac12}}_{\defines \Am_k}\underbrace{\LB\Qm_{k_a}-\Qm_{k_b}\RB}_{\defines \Bm_k}\RSB^2
\end{align}
where $\Am_k$ are Hermitian nonnegative definite and $\Bm_k$ are Hermitian. Let $\Am_k= \Um_k\Dm_k\Um_k\htp$ be the eigenvalue decomposition of $\Am_k$, where $\Um_k\in\CC^{n_k\times n_k}$ are unitary matrices and $\Dm_k=\diag\LB d_{k,1},\dots,d_{k,n_k}\RB \succeq 0$. Moreover, denote $\Zm_k = \Bm_k\Um_k = \LSB\zv_{k,1}\dots\zv_{k,n_k}\RSB$. Then,
\begin{align}\label{eq:traceresult}
 \frac1N\trace\LSB\Am_k\Bm_k\RSB^2 &= \frac1N\trace\Dm_k\Um\htp\Bm_k\Am_k\Bm_k\Um = \frac1N\trace\Dm_k\Zm_k\htp\Am_k\Zm_k=\frac1N\sum_{j=1}^{n_k} d_{k,j} \zv_{k,j}\htp\Am_k\zv_{k,j}\ge 0.
\end{align}
If $\Tm_k\succ 0$, or equivalently, if $\Tm_k$ is invertible, we have $\Am_k \succ 0$ and \eqref{eq:traceresult} holds with strict inequality for $\Qm_{k_a}\ne\Qm_{k_b}$. Thus, if any of the matrices $\Tm_k$ is invertible and $\Qm_{k_a}\ne\Qm_{k_b}$, we have 
\begin{align}
 \sum_{k=1}^K\frac1N\trace\LSB \Am_k\Bm_k\RSB^2 > 0
\end{align}
and hence $\frac{d^2 \bar{I}_N(\rho)}{d \lambda^2} < 0$. Thus, $\bar{I}_N(\rho)$ is strictly concave in the matrices $\Qm_k$.
Due to \eqref{eq:partderiv} it is then sufficient to maximize
\begin{align}
 \log\det\LB\Id_{n_k} + g_k\Tm_k^{\frac12}\Qm_k\Tm_k^{\frac12}\RB
\end{align}
with respect to $\Qm_k$ and with the constraint $\frac1{n_k}\trace\Qm_k\le P_k$. The solution to this problem is well-known \cite{COV06} and given by the water-filling solution stated in the theorem. However, since $g_k$ depends on $\Qm_k$, this solution must be computed iteratively by Algorithm~1.

\section{Proof of Theorem~\ref{th:sinr}: SINR of the MMSE detector}\label{proof:sinr}
Similar to the proof of Theorem~\ref{th:mutinfscatter}, let us consider the matrix model
\begin{align}
 \Hm_k = \frac{1}{\sqrt{n_k}}\Zm_k\Wm_{2,k}\Tm_k^{\frac12}\Qm_k^{\frac12}
\end{align}
where 
\begin{align}
 \Zm_k = \frac1{\sqrt{N_k}}\Rm_k^{\frac12}\Wm_{k,1}\Sm_k^{\frac12}.
\end{align}
Then,
\begin{align}
 \gamma_{k,j}^N = p_{k,j}t_{k,j} \frac1{n_k}\wv_{2,k,j}\htp\Zm_k\htp \LB \sum_{i=1}^K\Hm_i\Hm_i\htp  - p_{k,j}t_{k,j} \frac1{n_k}\Zm_k\wv_{2,k,j}\wv_{2,k,j}\Zm_k\htp+ \frac1{\rho}\Id_N\RB^{-1}\Zm_k\wv_{2,k,j}.
\end{align}
It was shown in the proof of Theorem~\ref{th:mutinfscatter} that, almost surely, $\limsup_N\lVert\Zm_k\Zm_k\htp \rVert<0$. From the Fubini theorem, Lemma~\ref{lem:trace} and Lemma~\ref{lem:rank1perturbation}, we therefore have 
\begin{align}
 \gamma_{k,j}^N - p_{k,j}t_{k,j}\frac1{n_k}\trace\Zm_k\Zm_k\htp\LB \sum_{i=1}^K \Hm_i\Hm_i\htp + \rho\Id_{n_k}\RB \xrightarrow[N\to\infty]{\text{a.s}}0.
\end{align}
Applying Theorem~\ref{th:mutinf}~$(i)$ leads to
\begin{align}
 \gamma_{k,j}^N - p_{k,j}t_{k,j}\frac1{n_k}\trace\Zm_k\Zm_k\htp\LB \sum_{i=1}^K \bar{e}_i\Zm_i\Zm_i\htp + \rho\Id_{n_k}\RB^{-1} \xrightarrow[N\to\infty]{\text{a.s}}0
\end{align}
 where $\bar{e}_i$ are given as the unique solutions to \eqref{eq:fundeqkron}. Notice now from \eqref{eq:fundeqkron} that \begin{align}e_k =\frac1{n_k}\trace\Zm_k\Zm_k\htp\LB \sum_{i=1}^K \bar{e}_i\Zm_i\Zm_i\htp + \rho\Id_{n_k}\RB^{-1}\end{align} and that $\max_k|e_k-g_k|\xrightarrow[N\to\infty]{\text{a.s}}0$ by \eqref{eq:convek}. This finally implies
\begin{align}
 \gamma_{k,j}^N - p_{k,j}t_{k,j} g_{k} \xrightarrow[N\to\infty]{\text{a.s}}0.
\end{align}

\section{Proof of Corollary~\ref{cor:rate}: Sum-rate with MMSE decoding }\label{proof:rate}
Part $(i)$ is a simple consequence of Theorem~\ref{th:sinr} and the continuous mapping theorem.

For Part $(ii)$, first notice that $R_N(\rho)\le I_N(\rho)$ and $\bar{R}_N(\rho)\le \bar{I}_N(\rho)$. Thus, 
\begin{align}
 \left|R_N(\rho) -\frac1N\sum_{k=1}^K\sum_{j=1}^{n_k}\log\LB 1 + p_{k,j}t_{k,j} g_{k} \RB \right| \le I_N(\rho) + \bar{I}_N(\rho) \defines \phi_N(\rho).
\end{align}
Since $\mathbb{E}\LSB\phi_N(\rho)\RSB<\infty$ by Theorem~\ref{th:mutinfscatter}~$(ii)$, it follows from dominated convergence arguments that
\begin{align}
 \mathbb{E}\LSB R_N(\rho)\RSB - \bar{R}_N(\rho) \xrightarrow[N\to\infty]{}0.
\end{align}

\section{Proof of Corollary~\ref{cor:rayprod}: Rayleigh product channel}\label{proof:rayprod}
Under the assumptions of the corollary, the fundamental equations in Theorem~\ref{th:fundequ} reduce to
\begin{align}\label{eq:1}
 \bar{g} &= \frac{1}{1+g}\\\label{eq:2}
g &= \frac SN \frac{\delta}{1+\bar{g}\delta}\\\label{eq:3}
\delta &= \frac1{K \frac{\bar{g} g}{\delta} + \frac SN\frac1\rho}
\end{align}
From \eqref{eq:1}, we have
\begin{align}\label{eq:4}
 g = \frac{1-\bar{g}}{\bar{g}}.
\end{align}
Solving \eqref{eq:2} for $\delta$ and replacing $g$ by \eqref{eq:4} yields
\begin{align}\label{eq:5}
 \delta = \frac{1-\bar{g}}{\bar{g}\LB\bar{g}+\frac SN - 1\RB}.
\end{align}
Solving \eqref{eq:3} for $\delta$ and replacing $g$ by \eqref{eq:4} leads to
\begin{align}\label{eq:6}
 \delta = \frac{1-K(1-\bar{g})}{\frac SN \frac1\rho}.
\end{align}
Equating \eqref{eq:5} and \eqref{eq:6} and rearranging the terms as a polynomial in $\bar{g}$ finally yields
\begin{align}\label{eq:7}
 \bar{g}^3 - \bar{g}^2\LB2-\frac SN - \frac1K\RB +\bar{g}\LB1-\frac SN - \frac1K + \frac S{NK}\LB 1+\frac1\rho\RB\RB - \frac{S}{NK}\frac1\rho = 0.
\end{align}
By Theorem~\ref{th:fundequ}, only one of the roots of this polynomial satisfies $\bar{g}, g, \delta >0$.
Now, \eqref{eq:4} implies $\bar{g}<1$, \eqref{eq:5} implies $\bar{g}>1-\frac SN$ \eqref{eq:6} implies $\bar{g}>1-\frac{1}{K}$. Hence $\bar{g}\in\LB1-\min\LSB\frac1K,\frac SN\RSB,1\RB$.

Similarly, $\bar{I}_N\LB\rho\RB$ reduces under the assumptions of the corollary to
\begin{align}
 \bar{I}_N\LB\rho\RB=  \log\LB1+\rho\frac{NK}{S}\frac{\bar{g} g}{\delta}\RB + \frac{KS}{N}\log\LB1+\bar{g}\delta\RB + K\log\LB1+g\RB - 2K\bar{g}g.
\end{align}
Replacing $\frac{g}{\delta}$ by $\bar{g}+\frac SN - 1$ in the first term, $\delta$ by \eqref{eq:5} in the second term, $g$ by \eqref{eq:4} in the third term and $\bar{g}g$ by $(1-\bar{g})$ in the last term leads to the desired result.

The simplification of Theorem~\ref{th:sinr} is immediate since $\bar{\gamma}^N_{kj}= p_{k,j}t_{k,j}g_k = \frac{1-\bar{g}}{\bar{g}}$ by \eqref{eq:4}.

\clearpage

\section{Matlab code related to the multihop AF MIMO relay channel}\label{sec:matlab}
\subsection{Code for Corollary~\ref{cor:mutinfrelay}: \texttt{corollary1.m}}\vspace{-25pt}
\begin{lstlisting}
function I_k = corollary1(k,K,a,rho,c)
% Compute deterministic equivalent I_k of the asymptotic mutual information
% Input parameters :
% k                             : denotes which function I_k to compute
% K                             : total number number of hops
% a   = [alpha_1,...,alpha_{K}] : vector containing the path loss factors
% rho = [rho_0,...,rho_{K-1}]   : vector containing the power budgets rho_k
% c   = [c_1,...,c_K]           : vector containing the matrix dimension ratios 
% Calculate asymptotic power normalization factors (Lemma 1)
b  = zeros(1,k);    
b(1) = rho(1);
for i=2:k
    b(i) = rho(i) / (1+a(i-1)*rho(i-1));
end
% Calculate capacity
I_k = 1/K*(theorem2(k,1,a,b,c) - theorem2(k,1,a,[0,b(2:end)],c));
end
\end{lstlisting}

\subsection{Code for Theorem~\ref{th:mutinfrelay}: \texttt{theorem2.m}}\vspace{-25pt}
\begin{lstlisting}
function J = theorem2(k,x,a,b,c)
% Recursively computes deterministic equivalent of J_k
% Input parameters :
% k                           : denotes which function J_k to compute
% x                           : argument of J_k
% a = [alpha_1,...,alpha_{k}] : vector containing the path loss factors
% b = [beta_0,...,beta_{k-1}] : power normalization factors
% c = [c_1,...,c_K]           : vector containing the matrix dimension ratios 
if (k==1) % J_1 is given in closed form
    [~,e] = theorem3(k-1,x,a,b,c);
    J     = c(k)*log(1 + x*a(k)*b(k)/(c(k)+e)) + log(1+e/c(k)) - e/(c(k)+e);
else      % J_k, k>1, must be computed recursively
    [~,e] = theorem3(k-1,x,a,b,c);
    J     = c(k)*theorem2(k-1,x*a(k)*b(k)/(c(k)+x*a(k)*b(k) + e),a,b,c)...
            + c(k)*log(1 + x*a(k)*b(k)/(c(k)+e)) + log(1 + e/c(k)) - e/(c(k)+e);
end
end
\end{lstlisting}

\subsection{Code for Theorem~\ref{th:mkek}: \texttt{theorem3.m}}\vspace{-25pt}
\begin{lstlisting}
function [m,e] = theorem3(k,x,a,b,c)
% Recursively computes the quantities e_k
% Input parameters :
% k                           : denotes which function e_k to compute
% x                           : argument of e_k
% a = [alpha_1,...,alpha_{k}] : vector containing the path loss factors
% b = [beta_0,...,beta_{k}]   : power normalization factors
% c = [c_1,...,c_K]           : vector containing the matrix dimension ratios 
if (k==0) % For k=0, e_o and m_0 are given in closed form
    e = -(x*a(1)*b(1)*(1-c(1))+c(1))/2 + sqrt((x*a(1)*b(1)*(1-c(1))+c(1))^2...
        + 4*x*a(1)*b(1)*c(1)^2)/2;
    m = c(1) / (a(1)*b(1)/(c(1)+e)+1/x) + (1-c(1))*x;
else      % For k>0, e_k is given as the solution to a fixed point equation
    e    = 0;
    eold = 1;
    while abs(e-eold)>1e-6  % the error tolerance 1e-6 can be changed 
        eold = e;
        e = c(k+1)*(c(k+1)+eold) - c(k+1)*(c(k+1)+eold)^2/(x*a(k+1)*b(k+1))...
            * theorem3(k-1,x*a(k+1)*b(k+1)/(c(k+1)+x*a(k+1)*b(k+1)+eold),a,b,c);    
    end
    m = x*c(k+1)/(c(k+1)+e);
end
end
\end{lstlisting}

\bibliography{IEEEabrv,bibliography,tutorial_RMT}

% Generated by IEEEtran.bst, version: 1.13 (2008/09/30)
\begin{thebibliography}{10}
\providecommand{\url}[1]{#1}
\csname url@samestyle\endcsname
\providecommand{\newblock}{\relax}
\providecommand{\bibinfo}[2]{#2}
\providecommand{\BIBentrySTDinterwordspacing}{\spaceskip=0pt\relax}
\providecommand{\BIBentryALTinterwordstretchfactor}{4}
\providecommand{\BIBentryALTinterwordspacing}{\spaceskip=\fontdimen2\font plus
\BIBentryALTinterwordstretchfactor\fontdimen3\font minus
  \fontdimen4\font\relax}
\providecommand{\BIBforeignlanguage}[2]{{%
\expandafter\ifx\csname l@#1\endcsname\relax
\typeout{** WARNING: IEEEtran.bst: No hyphenation pattern has been}%
\typeout{** loaded for the language `#1'. Using the pattern for}%
\typeout{** the default language instead.}%
\else
\language=\csname l@#1\endcsname
\fi
#2}}
\providecommand{\BIBdecl}{\relax}
\BIBdecl

\bibitem{TSE99}
D.~N.~C. Tse and S.~V. Hanly, ``{Linear multiuser receivers: effective
  interference, effective bandwidth and user capacity},'' \emph{{IEEE} Trans.
  Inf. Theory}, vol.~45, no.~2, pp. 641--657, Feb. 1999.

\bibitem{TUL04b}
A.~M. Tulino and S.~Verd\'u, ``{Random matrix theory and wireless
  communications},'' \emph{Foundations and Trends in Communications and
  Information Theory}, vol.~1, no.~1, 2004.

\bibitem{COUbook}
R.~Couillet and M.~Debbah, \emph{{Random matrix methods for wireless
  communications}}, 1st~ed.\hskip 1em plus 0.5em minus 0.4em\relax New York,
  NY, USA: Cambridge University Press, 2011.

\bibitem{TEL99}
I.~E. Telatar, ``{Capacity of multi-antenna Gaussian channels},''
  \emph{European Transactions on Telecommunications}, vol.~10, no.~6, pp.
  585--595, Feb. 1999.

\bibitem{HAC07}
W.~Hachem, P.~Loubaton, and J.~Najim, ``{Deterministic equivalents for certain
  functionals of large random matrices},'' \emph{Annals of Applied
  Probability}, vol.~17, no.~3, pp. 875--930, 2007.

\bibitem{DEB03b}
M.~Debbah, P.~Loubaton, and M.~{de Courville}, ``{The spectral efficiency of
  linear precoders},'' in \emph{(ITW'03)}, Paris, France, Mar. 2003, pp.
  90--93.

\bibitem{couillethoydis11}
\BIBentryALTinterwordspacing
R.~Couillet, J.~Hoydis, and M.~Debbah, ``Random beamforming over quasi-static
  and fading channels: A deterministic equivalent approach,'' \emph{{IEEE}
  Trans. Inf. Theory}, 2011, submitted. [Online]. Available:
  \url{http://arxiv.org/pdf/1011.3717v2}
\BIBentrySTDinterwordspacing

\bibitem{NGU08}
V.~K. Nguyen and J.~S. Evans, ``Multiuser transmit beamforming via regularized
  channel inversion: A large system analysis,'' in \emph{Proc. IEEE Global
  Communications Conference (Globecom’08)}, New Orleans, LO, US, Dec. 2008,
  pp. 1--4.

\bibitem{WAG10}
\BIBentryALTinterwordspacing
S.~Wagner, R.~Couillet, M.~Debbah, and D.~T.~M. Slock, ``{Large system analysis
  of linear precoding in MISO broadcast channels with limited feedback},''
  \emph{{IEEE} Trans. Inf. Theory}, 2011. [Online]. Available:
  \url{http://arxiv.org/abs/0906.3682}
\BIBentrySTDinterwordspacing

\bibitem{ZAK10}
R.~Zakhour and S.~Hanly, ``Large system analysis of base station cooperation on
  the downlink,'' in \emph{Proc. 48th Annual Allerton Conference on
  Communication, Control, and Computing}, Urbana-Champaign, IL, US, Oct. 2010,
  pp. 270 --277.

\bibitem{huh2010}
H.~Huh, G.~Caire, S.-H. Moon, and I.~Lee, ``Multi-cell {MIMO} downlink with
  fairness criteria: {T}he large system limit,'' in \emph{Proc. International
  Symposium on Information Theory Proceedings (ISIT'2010)}, Jun. 2010, pp.
  2058--2062.

\bibitem{HOY10}
J.~Hoydis, M.~Kobayashi, and M.~Debbah, ``Optimal channel training in uplink
  network {MIMO} systems,'' \emph{{IEEE} Trans. Signal Process.}, vol.~59,
  no.~6, Jun. 2011.

\bibitem{TUL04}
L.~Li, A.~M. Tulino, and S.~Verd\'u, ``{Design of reduced-rank MMSE multiuser
  detectors using random matrix methods},'' \emph{{IEEE} Trans. Inf. Theory},
  vol.~50, no.~6, pp. 986--1008, Jun. 2004.

\bibitem{COU09}
R.~Couillet, M.~Debbah, and J.~W. Silverstein, ``{A deterministic equivalent
  for the analysis of correlated MIMO multiple access channels},'' \emph{{IEEE}
  Trans. Inf. Theory}, vol.~57, no.~6, pp. 3493--3514, Jun. 2011.

\bibitem{DUP10}
\BIBentryALTinterwordspacing
F.~Dupuy and P.~Loubaton, ``{On the capacity achieving covariance matrix for
  frequency selective MIMO channels using the asymptotic approach},''
  \emph{{IEEE} Trans. Inf. Theory}, 2010. [Online]. Available:
  \url{http://arxiv.org/abs/1001.3102}
\BIBentrySTDinterwordspacing

\bibitem{HAC06}
W.~Hachem, O.~Khorunzhy, P.~Loubaton, J.~Najim, and L.~A. Pastur, ``{A new
  approach for capacity analysis of large dimensional multi-antenna
  channels},'' \emph{{IEEE} Trans. Inf. Theory}, vol.~54, no.~9, 2008.

\bibitem{MOU05}
A.~L. Moustakas and S.~H. Simon, ``{Random matrix theory of multi-antenna
  communications: the Rician channel},'' \emph{Journal of Physics A:
  Mathematical and General}, vol.~38, no.~49, pp. 10\,859--10\,872, Nov. 2005.

\bibitem{MOU07}
------, ``{On the outage capacity of correlated multiple-path MIMO channels},''
  \emph{{IEEE} Trans. Inf. Theory}, vol.~53, no.~11, pp. 3887--3903, 2007.

\bibitem{BIL08}
P.~Billingsley, \emph{{Probability and Measure}}, 3rd~ed.\hskip 1em plus 0.5em
  minus 0.4em\relax Hoboken, NJ: John Wiley and Sons, Inc., 1995.

\bibitem{BOR07}
S.~Borade, L.~Zheng, and R.~Gallager, ``{Amplify-and-forward in wireless relay
  networks: rate, diversity, and network size},'' \emph{{IEEE} Trans. Inf.
  Theory}, vol.~53, no.~10, pp. 3302--3318, 2007.

\bibitem{MAR10b}
I.~Maric, A.~Goldsmith, and M.~{M\'edard}, ``{Analog network coding in the
  high-SNR regime},'' in \emph{IEEE Wireless Network Coding Conference
  (WiNC'10)}, Boston, MA, USA, 2010, pp. 1--6.

\bibitem{FAW09}
N.~Fawaz, K.~Zarifi, M.~Debbah, and D.~Gesbert, ``{Asymptotic capacity and
  optimal precoding in MIMO multi-hop relay networks},'' \emph{{IEEE} Trans.
  Inf. Theory}, vol.~57, no.~4, pp. 2050--2069, 2011.

\bibitem{gesbert02}
D.~Gesbert, H.~Bolcskei, D.~A. Gore, and A.~J. Paulraj, ``Outdoor {MIMO}
  wireless channels: models and performance prediction,'' \emph{{IEEE} Trans.
  Commun.}, vol.~50, no.~12, pp. 1926--1934, Dec. 2002.

\bibitem{muller01}
R.~R. M\"{u}ller and H.~Hofstetter, ``Confirmation of random matrix model for
  the antenna array channel by indoor measurements,'' in \emph{Proc. IEEE Int.
  Symp. Antennas and Propagation Society}, vol.~1, 2001, pp. 472--475.

\bibitem{DUM10}
J.~Dumont, W.~Hachem, S.~Lasaulce, P.~Loubaton, and J.~Najim, ``{On the
  capacity achieving covariance matrix for Rician MIMO channels: an asymptotic
  approach},'' \emph{{IEEE} Trans. Inf. Theory}, vol.~56, no.~3, pp.
  1048--1069, 2010.

\bibitem{tucci11}
G.~H. Tucci, ``Spectral analysis of the amplify and forward relay network as
  the number of relay layers increases,'' in \emph{Proc. 8th International
  Symposium on Wireless Communication Systems (ISWCS'11)}, Aachen, Germany,
  Nov. 2011.

\bibitem{leveque07}
S.-P. Yeh and O.~Leveque, ``Asymptotic capacity of multi-level
  amplify-and-forward relay networks,'' in \emph{Proc. IEEE International
  Symposium on Information Theory (ISIT'07)}, Jun. 2007, pp. 1436--1440.

\bibitem{morgenshtern06}
V.~I. Morgenshtern and H.~Bölcskei, ``Random matrix analysis of large relay
  networks,'' in \emph{Proc. 44th Allerton Annual Conference on Communications,
  Control and Computing}, Urbana-Champaign, IL, US, Sep. 2006, pp. 106--112.

\bibitem{wagner08}
J.~Wagner, B.~Rankov, and A.~Wittneben, ``Large n analysis of
  amplify-and-forward {MIMO} relay channels with correlated {Rayleigh}
  fading,'' \emph{{IEEE} Trans. Inf. Theory}, vol.~54, no.~12, pp. 5735--5746,
  Dec. 2008.

\bibitem{jin10}
S.~Jin, M.~R. McKay, C.~Zhong, and K.-K. Wong, ``Ergodic capacity analysis of
  amplify-and-forward {MIMO} dual-hop systems,'' \emph{{IEEE} Trans. Inf.
  Theory}, vol.~56, no.~5, pp. 2204--2224, May 2010.

\bibitem{vdW}
A.~W. van~der Vaart, \emph{Asymptotic Statistics (Cambridge Series in
  Statistical and Probabilistic Mathematics)}.\hskip 1em plus 0.5em minus
  0.4em\relax Cambridge University Press, New York, 1998.

\bibitem{muller02}
R.~R. M\"{u}ller, ``A random matrix model of communication via antenna
  arrays,'' \emph{{IEEE} Trans. Inf. Theory}, vol.~48, no.~9, pp. 2495--2506,
  Sep. 2002.

\bibitem{chizik02}
D.~Chizhik, G.~J. Foschini, M.~J. Gans, and R.~A. Valenzuela, ``Keyholes,
  correlations, and capacities of multielement transmit and receive antennas,''
  \emph{{IEEE} Trans. Wireless Commun.}, vol.~1, no.~2, pp. 361--368, Apr.
  2002.

\bibitem{almers06}
P.~Almers, F.~Tufvesson, and A.~F. Molisch, ``Keyhole effect in {MIMO} wireless
  channels: {M}easurements and theory,'' \emph{{IEEE} Trans. Wireless Commun.},
  vol.~5, no.~12, pp. 3596--3604, Dec. 2006.

\bibitem{shin03}
H.~Shin and J.~H. Lee, ``Capacity of multiple-antenna fading channels: spatial
  fading correlation, double scattering, and keyhole,'' \emph{{IEEE} Trans.
  Inf. Theory}, vol.~49, no.~10, pp. 2636--2647, Oct. 2003.

\bibitem{levin06}
G.~Levin and S.~Loyka, ``Multi-keyhole {MIMO} channels: asymptotic analysis of
  outage capacity,'' in \emph{Proc. IEEE International Symposium on Information
  Theory (ISIT'06)}, Seattle, Washington, US, Jul. 2006, pp. 1305--1309.

\bibitem{shin08}
H.~Shin and M.~Z. Win, ``{MIMO} diversity in the presence of double
  scattering,'' \emph{{IEEE} Trans. Inf. Theory}, vol.~54, no.~7, pp.
  2976--2996, Jul. 2008.

\bibitem{yang11}
S.~Yang and J.-C. Belfiore, ``Diversity-multiplexing tradeoff of double
  scattering {MIMO} channels,'' \emph{{IEEE} Trans. Inf. Theory}, vol.~57,
  no.~4, pp. 2027--2034, Apr. 2011.

\bibitem{jin08}
S.~Jin, M.~R. McKay, K.-K. Wong, and X.~Gao, ``Transmit beamforming in
  {R}ayleigh product {MIMO} channels: capacity and performance analysis,''
  \emph{{IEEE} Trans. Signal Process.}, vol.~56, no.~10, pp. 5204--5221, Oct.
  2008.

\bibitem{li10}
X.~Li, S.~Jin, X.~Gao, and M.~R. McKay, ``Capacity bounds and low complexity
  transceiver design for double-scattering {MIMO} multiple access channels,''
  \emph{{IEEE} Trans. Signal Process.}, vol.~58, no.~5, pp. 2809--2822, May
  2010.

\bibitem{COV06}
T.~Cover and J.~A. Thomas, \emph{{Elements of Information Theory}},
  2nd~ed.\hskip 1em plus 0.5em minus 0.4em\relax New York, NY, USA: John Wiley
  and Sons, Inc., 2006.

\bibitem{kaybook}
S.~M. Kay, \emph{Fundamentals of Statistical Signal Processing: Estimation
  Theory}.\hskip 1em plus 0.5em minus 0.4em\relax Prentice-Hall, Inc. Upper
  Saddle River, NJ, USA, 1993.

\bibitem{GES02}
D.~Gesbert, T.~Ekman, and N.~Christophersen, ``Capacity limits of dense
  palm-sized {MIMO} arrays,'' in \emph{Proc. Global Telecommunications
  Conference (GLOBECOM '02)}, vol.~2, Taipei, Taiwan, Nov. 2002, pp.
  1187--1191.

\bibitem{tsebook}
D.~Tse and P.~Viswanath, \emph{Fundamentals of Wireless Communications}.\hskip
  1em plus 0.5em minus 0.4em\relax New York, NY, USA: Cambridge University
  Press, 2005.

\bibitem{KAM10}
W.~H. A.~Kammoun, M.~Kharouf and J.~Najim, ``A central limit theorem for the
  {SINR} at the {LMMSE} estimator output for large dimensional systems,''
  \emph{{IEEE} Trans. Inf. Theory}, vol.~55, pp. 5048--5063, Nov. 2009.

\bibitem{SIL98}
Z.~D. Bai and J.~W. Silverstein, ``{No eigenvalues outside the support of the
  limiting spectral distribution of large dimensional sample covariance
  matrices},'' \emph{The Annals of Probability}, vol.~26, no.~1, pp. 316--345,
  Jan. 1998.

\bibitem{SIL95}
J.~W. Silverstein and Z.~D. Bai, ``{On the empirical distribution of
  eigenvalues of a class of large dimensional random matrices},'' \emph{Journal
  of Multivariate Analysis}, vol.~54, no.~2, pp. 175--192, 1995.

\bibitem{yates95}
R.~Yates, ``A framework for uplink power control in cellular radio systems,''
  \emph{{IEEE} J. Sel. Areas Commun.}, vol.~13, no.~7, pp. 1341--1347, Sep.
  1995.

\bibitem{titchmarsh}
E.~C. Titchmarsh, \emph{The Theory of Functions}.\hskip 1em plus 0.5em minus
  0.4em\relax Oxford University Press, London, Aug. 1939.

\bibitem{boydcvx}
S.~Boyd and L.~Vandenberghe, \emph{{Convex Optimization}}.\hskip 1em plus 0.5em
  minus 0.4em\relax Cambridge University Press, 2004.

\end{thebibliography}

\end{document}